\documentclass[journal=jacsat,manuscript=article]{achemso}

\usepackage[version=3]{mhchem} 
\usepackage{graphicx}
\usepackage{bm}
\usepackage{lscape}
\usepackage{amsmath}
\usepackage{amssymb}
\usepackage{braket}
\usepackage{longtable}
\usepackage{array}
\usepackage{natbib}
\usepackage{xcolor} 
\usepackage{epstopdf}
\SectionNumbersOn

\author{Jun Jiang}
\affiliation[Lawrence Livermore National Laboratory]
{Center for Accelerator Mass Spectrometry, Lawrence Livermore National Laboratory, Livermore, California 94550, USA}
\author{Hong-Zhou Ye}
\affiliation[Massachusetts Institute of Technology]
{Department of Chemistry, Massachusetts Institute of Technology, Cambridge, Massachusetts 02139, USA}
\altaffiliation{Current address: Department of Chemistry, Columbia University, New York, New York 10027 USA}
\author{Klaas Nauta}
\affiliation[UNSW]
{School of Chemistry, UNSW, Sydney, NSW 2052, Australia}
\author{Troy Van Voorhis}
\affiliation[Massachusetts Institute of Technology]
{Department of Chemistry, Massachusetts Institute of Technology, Cambridge, Massachusetts 02139, USA}
\author{Timothy W. Schmidt}
\affiliation[UNSW]
{School of Chemistry, UNSW, Sydney, NSW 2052, Australia}
\author{Robert W. Field}
\email{rwfield@mit.edu}
\phone{617-253-1489}
\affiliation[Massachusetts Institute of Technology]
{Department of Chemistry, Massachusetts Institute of Technology, Cambridge, Massachusetts 02139, USA}

\title
  {Diabatic valence-hole states in the C$_2$ molecule:\\
 ``Putting Humpty Dumpty together again''}

\begin{document}

\begin{tocentry}
\center
\includegraphics{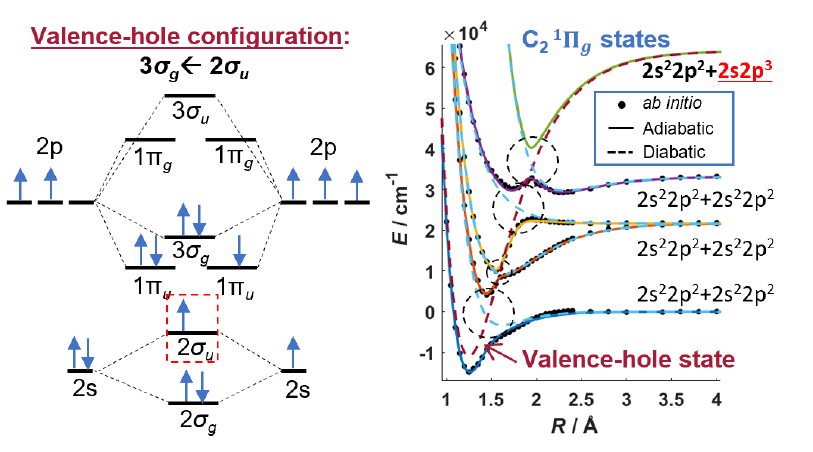}

\end{tocentry}

\newpage
\begin{abstract}
Despite the long history of spectroscopic studies of the C$_2$ molecule, fundamental questions about its chemical bonding are still being hotly debated. The complex electronic structure of C$_2$ is a consequence of its dense manifold of near-degenerate, low-lying electronic states. A global multi-state diabatic model is proposed here to disentangle the numerous configuration interactions within four symmetry manifolds of C$_2$ ($^{1}\Pi_g$, $^{3}\Pi_g$, $^{1}\Sigma_u^+$, and $^{3}\Sigma_u^+$). The key concept of our model is the existence of two ``valence-hole'' configurations, $2\sigma_g^22\sigma_u^11\pi_{u}^33\sigma_g^2$ for $^{1,3}\Pi_g$ states and $2\sigma_g^22\sigma_u^11\pi_{u}^43\sigma_g^1$ for $^{1,3}\Sigma_u^+$ states that derive from $3\sigma_g\leftarrow2\sigma_u$ electron promotion. The lowest-energy state from each of the four C$_2$ symmetry species is dominated by this type of valence-hole configuration at its equilibrium internuclear separation. As a result of their large binding energy (nominal bond order of 3) and correlation with the 2s$^2$2p$^2$+2s2p$^3$ separated-atom configurations, the presence of these valence-hole configurations has a profound impact on the $global$ electronic structure and unimolecular dynamics of C$_2$.
\end{abstract}

\newpage
\section{Introduction}
\label{sec:intro}

The C$_2$ molecule is often treated as a classic example in introductory quantum mechanics courses to illustrate the principles of molecular orbital (MO) theory in the linear combination of atomic orbitals (LCAO) approach. Despite the simple chemical composition of C$_2$, its electronic structure is surprisingly complicated~\cite{gulania2019eom}. As a result of the near-degeneracy among multiple electron configurations, the majority of the low-lying C$_2$ electronic states are poorly described by a single dominant electron configuration~\cite{hirsch1980non,Chabalowski1981,Chabalowski1983}. Even its electronic ground state ($X^1\Sigma_g^+$), which is the subject of ongoing debates about the chemical bonding nature of C$_2$~\cite{shaik2012quadruple,shaik2013one,frenking2013critical,danovich2013response,laws2019dicarbon}, is described by two dominant configurations, $0.828 \left |2\sigma_g^22\sigma_u^21\pi_{u}^4\right |-0.324\left |2\sigma_g^21\pi_{u}^43\sigma_g^2\right |+...$, at its equilibrium internuclear separation ($R_e$)~\cite{shaik2012quadruple}. In comparison, the electronic ground states of most of the other C/N/O diatomic molecules as well as larger ``relatives'' of C$_2$ (e.g. acetylene)~\cite{shaik2013one} are dominated by a single configuration. The multi-reference character of C$_2$ undercuts the utility of MO-theory concepts, such as bond order and dissociation correlation diagrams.

\begin{figure}
\center
\includegraphics[width=2.5in]{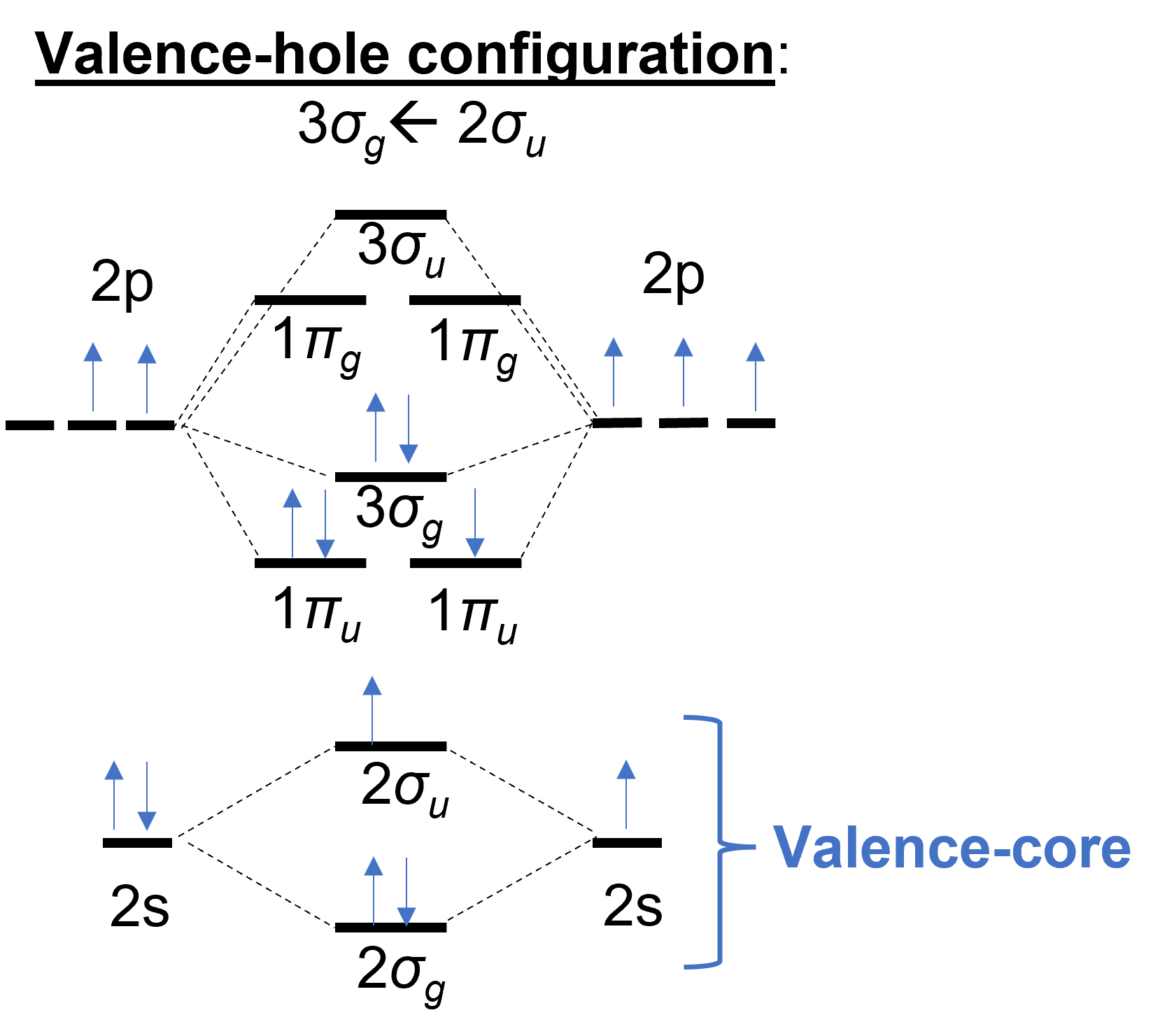}
\caption{The MO diagram for a valence-hole electron configuration. The long-range correlations between the atomic and molecular orbitals are given by the dotted lines.}
\label{fig:valence_hole_configuration}
\end{figure}

In this paper, we present a diabatic interaction model for the low-lying $^1\Pi_g$ states of C$_2$. This model can be extended to three other symmetry species ($^{1,3}\Sigma_u^+$ and $^3\Pi_g$). We hope that our work will create a new framework for discussions of the electronic structure of C$_2$. Our model for the $^1\Pi_g$ states is motivated by observations of unusual energy level patterns in the $C^1\Pi_g$ state, which is the upper state of the Deslandres-d'Azambujia band system ($C^1\Pi_g-A^1\Pi_u$)~\cite{DA1905,Dieke1930,Herzberg1940,Phillips1950,Messerle1967,Antic1985}. A diabatic fit model for the $C$ state allows us to account for these strange level patterns, as well as to conclusively invalidate the existence of the Messerle-Krauss band system ($C'^1\Pi_g-A^1\Pi_u$)~\cite{Messerle1967}. By treating the configuration mixing as a result of diabatic interactions, the usefulness of simple concepts of bond order and correlation diagrams is restored. Most importantly, the use of the diabatic picture allows us to uncover the specialness of the ``valence-hole'' configuration in C$_2$, e.g., $2\sigma_g^22\sigma_u^11\pi_{u}^33\sigma_g^2$, which has a triply-occupied valence-core (i.e.~$2\sigma_g^22\sigma_u^1$) and no electron in either the $1\pi_g$ or $3\sigma_u$ anti-bonding orbitals (Fig.~\ref{fig:valence_hole_configuration}). This type of valence-hole configuration that derives from $3\sigma_g\leftarrow2\sigma_u$ electron promotion exists in four symmetry manifolds of C$_2$ ($^{1,3}\Pi_g$ and $^{1,3}\Sigma_u^+$). In all four cases, the valence-hole configuration is the dominant configuration ($>$70$\%$ of the total character) at $R_e$ ($\sim$1.25\,$\textup{\AA}$) of the lowest energy state of each specified symmetry ($C^1\Pi_g$, $d^3\Pi_g$, $D^1\Sigma_u^+$, and $c^3\Sigma_u^+$)~\cite{hirsch1980non,Chabalowski1981,Chabalowski1983}.

The valence-hole configuration is a hitherto neglected feature in the $global$ electronic structure of C$_2$. As a result of the intrinsically large binding energy of the valence-hole states (nominal bond order of 3) and correlation with an atomic fragment channel with a single 2p$\leftarrow$2s electron promotion (``hole'' in the valence-core), the presence of these valence-hole configurations leads to large and systematic disruptions of the $global$ electronic structure of C$_2$.

\section{Motivations}

\begin{figure}
\center
\includegraphics[width=5 in]{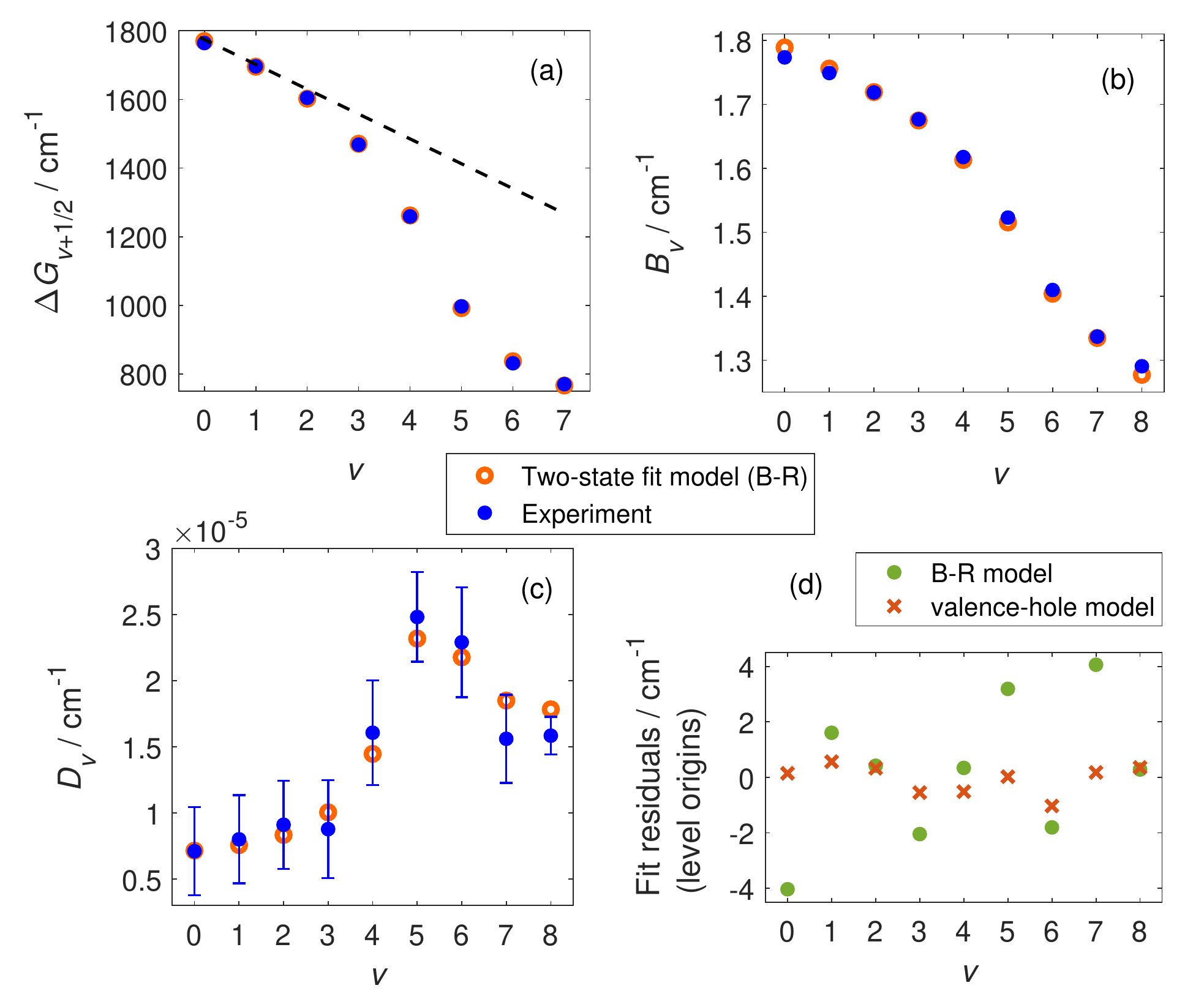}
\caption{The vibration-rotation constants of the C$_2$ $C^1\Pi_g$ state. (a) Vibrational energy spacings, $\Delta G_{v+1/2}$ (b) Rotational constants, $B_v$ (c) Centrifugal distortion constants, $D_v$ (d) Fit residuals for the vibrational level origins from the B-R model (green dots, Table~\ref{tab:fit}) and valence-hole two-state model (orange crosses, Table~S1 of the supplementary text). The error bars in (c) are $2\sigma$ uncertainties of the experimental values.}
\label{fig:exp_vs_fit}
\end{figure}

In our implementation of H-atom fluorescence action spectroscopy detection of predissociated S$_1$ levels of acetylene~\cite{jiang2018probing}, we discovered that one-color, resonance-enhanced (S$_1$$\leftarrow$S$_0$), multi-photon dissociation of acetylene leads to efficient production of the C$_2$ $C^1\Pi_g$ state~\cite{jiang2019one,jiang2020one}. In our analysis of the dispersed fluorescence spectra of the photofragments, we were struck by the highly unusual $C$-state vibration-rotation structure. In Figs.~\ref{fig:exp_vs_fit}a-\ref{fig:exp_vs_fit}c, the vibrational energy spacings ($\Delta G_{v+1/2}$), rotational constants ($B_v$), and centrifugal distortion constants ($D_v$), are shown, respectively, as a function of the $C$-state vibrational quantum number, $v$. The large curvature in the $v$-dependence of these vibration-rotation constants is a clear indication of the presence of strong, systematic perturbations. The early onset of these large curvatures, at $v\sim3$, which is $>$1\,eV below the dissociation limit, implies a strong deviation of the $C$-state potential energy curve (PEC) from a Morse-like potential.

\begin{figure}[t]
\center
\includegraphics[width=5.5 in]{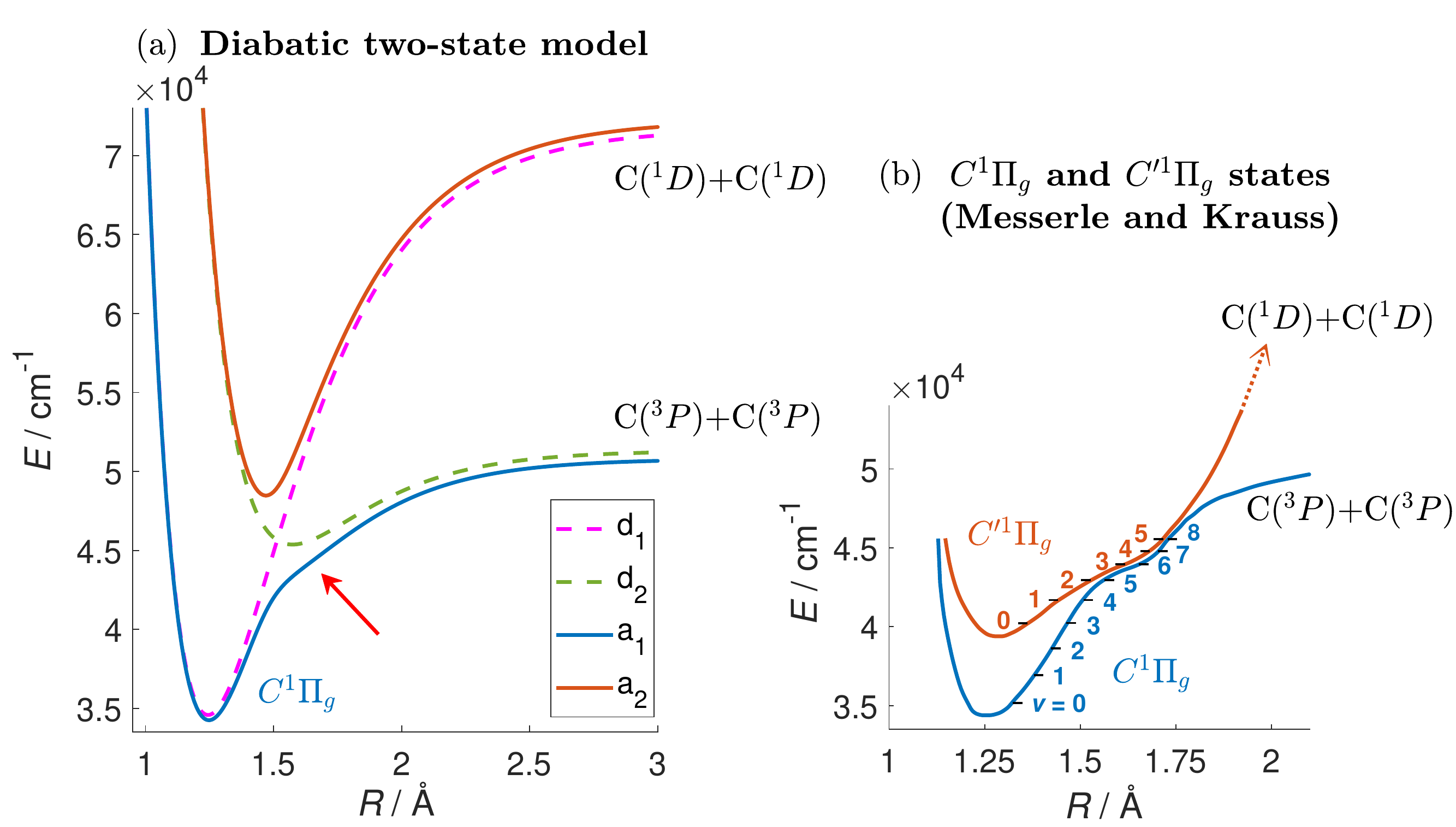}
\caption{The $^1\Pi_g$ potential curves. (a) The diabatic two-state model for the $C^1\Pi_g$ state is inspired by Ballik and Ramsay~\cite{Ballik1963}. The lower adiabat (a$_1$) corresponds to the $C^1\Pi_g$ state. (b) The $C^1\Pi_g$ and $C'^1\Pi_g$ states (according to Messerle and Krauss)~\cite{Messerle1967}. Edited, based on Fig. 4 of Ref.~\citenum{Messerle1967}.}
\label{fig:PECs}
\end{figure}

Ballik and Ramsay~\cite{Ballik1963} made the insightful comment that, based on the trend of $\Delta G_{v+1/2}$ at low $v$ (the dashed line in Fig.~\ref{fig:exp_vs_fit}a), the zeroth-order $C$ state ``wants to dissociate'' into the C($^1D$)+C($^1D$) limit, which is the second lowest possible dissociation channel for a $^1\Pi_g$ state~\cite{Ballik1963,herzberg1950molecular}. As $R$ increases, the zeroth-order $C$-state PEC inevitably crosses another $^1\Pi_g$-state PEC that dissociates into the lowest C($^3P$)+C($^3P$) channel. Due to the non-crossing rule, the $adiabatic$ $C$-state PEC is forced to bend downward toward C($^3P$)+C($^3P$). This $diabatic$ two-state interaction model for the C$_2$ $C$ state is illustrated in Fig.~\ref{fig:PECs}a. At a given $R$, the two adiabatic states (a$_1$ and a$_2$) are related to the diabatic states (d$_1$ and d$_2$) by the following representation,
\begin{equation} \label{mat}
	\begin{pmatrix}
 		E_1^d (R)\,\, &\,\, H_{12}^{el}(R) \\ 
		H_{12}^{el}(R)\,\, &\,\, E_2^d (R)
	\end{pmatrix},
\end{equation}
where $E_1^d(R)$ and $E_2^d(R)$ are the diabatic potential energies, and $H_{12}^{el}(R)$ expresses the electrostatic ($e^2/r_{ij}$) interaction between the two diabats. The two adiabatic PECs are obtained by diagonalizing Eq.~\ref{mat} as a function of $R$. Unlike an adiabatic state, the electronic character of a diabatic state is not a strong function of $R$~\cite{Bob2004}.

The diabatic interaction picture provides a simple, physically intuitive explanation for all of the strangenesses in the $C$-state vibration-rotation structure. The two-state model, as illustrated in Fig.~\ref{fig:PECs}a, is the Ballik-Ramsay (B-R) model. In our implementation of a B-R model (see Sections~\ref{sec:two-state} for details), the d$_1$ and d$_2$ PECs are both modeled as Morse potentials. The magnitude of the electrostatic interaction matrix element, $H_{12}^{el}$, is assumed to be $R$-independent. The large curvatures in the $v$-dependence of the $C$-state vibration-rotation constants are directly related to the sudden change of slope of its PEC, indicated by the arrow in Fig.~\ref{fig:PECs}a. The flattening of the $C$-state PEC is caused by a low-energy curve-crossing between two strongly-interacting diabats. This flattening of the PEC briefly mimics bond dissociation, and leads to the rapidly decreasing $\Delta G_{v+1/2}$ above $v=3$.

\begin{figure} [t]
\center
\includegraphics[width=3.3 in]{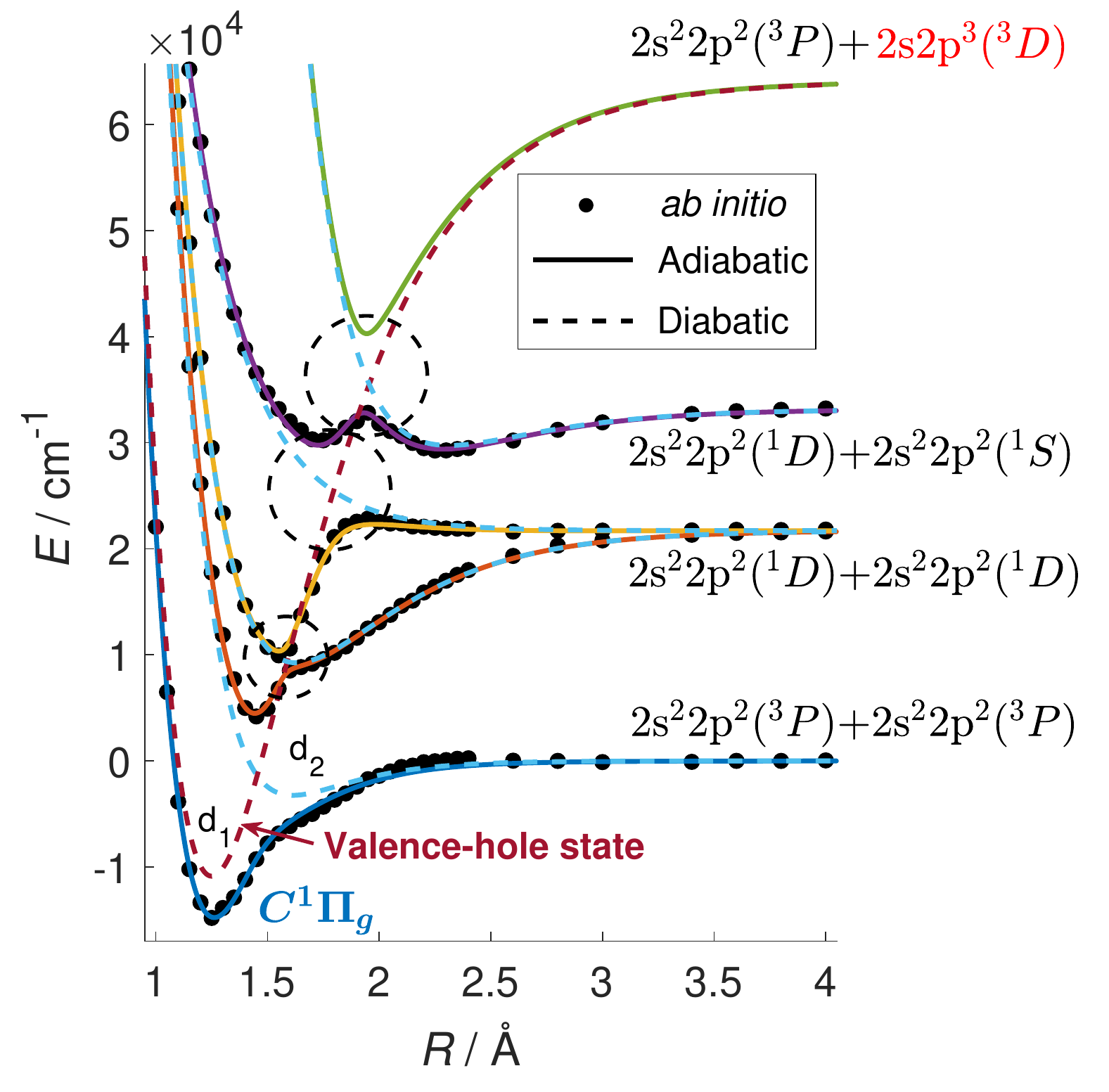}
\caption{The PECs for the lowest-energy $^1\Pi_g$ states from the $ab$ $initio$ calculation and our global five-state diabatic interaction model (see Section~\ref{sec:theory} for details). The $ab$ $initio$ potentials are calculated by the density matrix renormalization group (DMRG) method. The diabatic state labeled as d$_2$ in (a) corresponds to the same d$_2$ state in Fig.~\ref{fig:PECs}a. Note that the DMRG calculation provides only four roots at each $R$ value. Our global diabatization scheme for the  $^{1}\Pi_g$ states requires a five-state interaction. The qualitatively correct shape for the fifth adiabatic curve in our global interacting diabats model is confirmed by the results from a complete active space self-consistence field (CASSCF) level calculation, which includes this missing fifth root (see Fig.~S4a of the supplementary text).}
\label{fig:valence_hole}
\end{figure}

While the B-R model successfully reproduces most of the key spectroscopic anomalies of the $C^1\Pi_g$ state (Figs.~\ref{fig:exp_vs_fit}a-\ref{fig:exp_vs_fit}c), the fit residuals of the model ($\sim$2 cm$^{-1}$ on average) are much larger than the experimental uncertainties ($\sim$0.2 cm$^{-1}$)~\cite{Dieke1930,Herzberg1940,Phillips1950,Messerle1967,mckemmish2020update}. Upon inspection of the $ab$ $initio$ potentials for the lowest four $^1\Pi_g$ states (Fig.~\ref{fig:valence_hole}), we note that a crucial error was made in the assumption about the dissociation limit of the d$_1$ state in the B-R model. As indicated by the three dashed circles in Fig.~\ref{fig:valence_hole}, there are $three$ additional avoided-crossings, one on each of the 2$^{\textup{nd}}$, 3$^{\textup{rd}}$, and 4$^{\textup{th}}$ PECs. $These$ $curve$-$crossings$ $are$ $all$ $related$. Based on the shapes and locations of these avoided crossings, it appears that the d$_1$ state dissociates into a fragment channel at an energy above C($^1D$)+C($^1S$), which is the highest-energy dissociation channel for states of $^1\Pi_g$ symmetry in which both carbon atoms are in the 2s$^2$2p$^2$ configuration~\cite{Ballik1963,herzberg1950molecular}. The four $ab$ $initio$ PECs in Fig.~\ref{fig:valence_hole} are well reproduced using a multi-state diabatization scheme that assumes curve-crossings of a single, deeply-bound, diabatic state (d$_1$, dashed brown) with four other diabats (dashed light-blue). In this model, the d$_1$ state dissociates into a high-energy, 2s$^2$2p$^2$($^3P$)+$2s2p^3$($^3D$) channel, with one 2p$\leftarrow$2s electron promotion on one of the carbon atoms. An effective two-state (d$_1$ and d$_2$) fit model, which assumes this higher-energy dissociation limit for the d$_1$ state, significantly improves the numerical accuracy of the model ($\sim$0.4 cm$^{-1}$ average residual, see Fig.~\ref{fig:exp_vs_fit}d). We will interpret the exceptional stability of the d$_1$ diabat relative to its dissociation products using the concept of the ``valence-hole'' configuration. This is the key to our $global$ deperturbation of the diabatic interactions among the lowest five $^1\Pi_g$ states. This global diabatization scheme, based on the valence-hole concept, is extended to model the electronic structure of the $^{3}\Pi_g$, $^{1}\Sigma_u^+$, and $^{3}\Sigma_u^+$ states of C$_2$.

\section{Methods}
\label{sec:theory}

\subsection{Two-state fit model for the $C^1\Pi_g$ state}
\label{sec:two-state}

The Hamiltonian for the two-state model is constructed in the diabatic basis. The diagonal matrix elements are the zeroth-order electronic-vibration-rotation energies of the two diabats (d$_1$ and d$_2$ in Fig.~3a) at a specified $J$. The off-diagonal, electrostatic interaction matrix elements between ro-vibronic levels ($v$, $J$) of the two diabats are given by~\cite{Bob2004}
\begin{equation} \label{eq:He}
H_{1,v_1,J;2,v_2,J}=\bra{\Phi_1^d\xi_{v_1}^d J} H^{el} \ket{\Phi_2^d\xi_{v_2}^d J},
\end{equation}
where the diabatic ro-vibronic basis state, $\ket{\Phi_i^d\xi_{v_i}^d J}$, is expressed explicitly as a direct product of the electronic ($\Phi_i^d$), vibrational ($\xi_{v_i}^d$), and rotational ($J$) wavefunctions.

In the B-R two-state model, the two diabatic PECs are both modeled by a Morse potential,
\begin{equation} \label{eq:Morse}
T_e+D [1-e^{-\beta(R-R_e)}]^2,
\end{equation}
where $T_e$, $D$, $\beta$, and $R_e$ are the usual Morse potential parameters. The electronic part of the off-diagonal electrostatic interaction matrix elements (Eq.~\ref{eq:He}) is assumed to be independent of $R$,
\begin{equation} \label{eq:He_no_R}
\bra{\Phi_1^d}H^{el}\ket{\Phi_2^d}=H_{12}^{el}(R)=H_e.
\end{equation}
Under this assumption (which is discussed later in this section), the off-diagonal matrix elements can be simplified to 
\begin{equation} \label{eq:He_sim}
H_{1,v_1,J;2,v_2,J}=H_e \left \langle \xi_{v_1}^d J |  \xi_{v_2}^d J  \right \rangle_R,
\end{equation}
where $\left \langle \xi_{v_1}^d J |  \xi_{v_2}^d J  \right \rangle_R$ is the vibrational overlap integral between two diabatic ro-vibronic basis states. 

With the addition of the term, $\frac{\hbar^2}{2\mu R^2}[J(J+1)-\Omega^2]$, to the two diabatic PECs, where $\Omega=1$ for the $^1\Pi$ state, the energies and the vibrational wavefunctions of the two diabats are obtained using the Discrete Variable Representation (DVR) method~\cite{colbert1992novel}, with an evenly-spaced grid of $R$ values ($\Delta R=0.01\,\text{\AA}$) for $0.8\,\text{\AA}\leq R \leq R_{\mathrm{max}}=7\,\text{\AA}$. The off-diagonal matrix elements, Eq.~\ref{eq:He_sim}, are obtained by numerical integration over $R$ of the product of the DVR vibrational wavefunctions. The eigenvalues of the Hamiltonian are found to converge well, with respect to $R_{\mathrm{max}}$, $\Delta R$, and the number of basis states for the two diabats (100 basis states for d$_1$ and 500 basis states for d$_2$).

The assumption of an $R$-independent electronic matrix element (i.e. $H_{12}^{el}(R)=H_e$) does not significantly reduce the numerical accuracy of the two-state fit model. The $H_{1,v_1,J;2,v_2,J}$ matrix elements (Eq.~\ref{eq:He}) accumulate most of their amplitudes near the $R$-value of the intersection of the two diabats~\cite{Bob2004}. Given that the $R$-dependence of $H_{12}^{el}$ is typically weak, these electrostatic interaction matrix elements are more sensitive to the magnitude of $H_{12}^{el}$ at the diabatic curve-crossing region than the exact $R$-dependent form of $H_{12}^{el}$. As discussed in Section S2 of the supplementary text, use of qualitatively different $R$-dependent forms of the $H_{12}^{el}$ matrix element (constant, linearly-increasing, linearly-decreasing, and exponential-decay) leads to only modest changes in the numerical accuracy of the valence-hole two-state model. The average fit residuals range from 0.3 cm$^{-1}$ to 0.5 cm$^{-1}$ for different choices of $H_{12}^{el}(R)$. In addition, all of the derived $H_{12}^{el}$ matrix elements have approximately the same magnitude at the curve-crossing $R$-value of the two diabats, independent of the exact functional form of $H_{12}^{el}(R)$. 

Our two-state fit model for the $C^1\Pi_g$ state neglects the effects from both $\Lambda$-type doubling and nuclear-spin statistics. The $C$-state term values that are used as inputs to the fit model are obtained from rotational fits to the experimental observations, with the following expression for the effective term values,
\begin{equation} \label{rotformula}
F_v(J)=T_v+B_v[J(J+1)-1]-D_v[J(J+1)-1]^2.
\end{equation}
The two-state fit model is obtained by fitting the eigenvalues of the two-state Hamiltonian to the effective $C$-state term values at $J=1,5,10,15,$ and 20 of the $v=0-8$ levels. The deviations of these effective term values from the observed $C$-state level energies (which are affected by $\Lambda$-doubling) are typically $<$0.5 cm$^{-1}$ for $J\leq20$. The use of five selected $J$-levels in the two-state fit, instead of the full $J=1-20$ range, does not affect the validity of the fit model, reflecting the weak $J$-dependence for the electrostatic interaction matrix elements (Eq.~\ref{eq:He})~\cite{Bob2004}. In addition, the use of only five selected $J$-levels significantly increases the computational efficiency during the non-linear fit.

\subsection{Quantum chemical calculation}
\label{sec:ab_initio}

The $ab$ $initio$ potentials for the $^1\Pi_g$ states in Fig.~\ref{fig:valence_hole} are obtained using the density matrix renormalization group (DMRG) method with the cc-pVTZ basis. The calculation is implemented in the BLOCK code~\cite{chan2002highly,chan2004algorithm,ghosh2008orbital,sharma2012spin,olivares2015ab} and the PySCF software package~\cite{sun2018pyscf}. The DMRG potential curves for the lowest two $^1\Pi_g$ states agree well with the $ab$ $initio$ results reported in Ref.~\citenum{schmidt2021spectroscopy}, which are obtained using the multi-reference configuration interaction (MRCI) method. 

The $ab$ $initio$ potentials for the $^3\Pi_g$ states in Fig.~\ref{fig:valence_hole_3Pig}a are obtained using the MRCI method with the aug-cc-pCVQZ basis. The reference space is computed with a full-valence complete active space self-consistence field (CASSCF) calculation of the lowest $^3\Pi_g$ state. The MRCI wavefunctions contain all single and double excitations from the CASSCF reference wavefunction, including excitations from the 1s core. The calculation is implemented using the MOLPRO program~\cite{werner2020molpro}. Details for the derivation of the non-adiabatic interaction matrix elements from the MRCI calculation have been described in Ref.~\citenum{krechkivska2017first}. In Fig.~\ref{fig:valence_hole_3Pig}b, with proper choices for the signs of the MRCI wavefunctions, all six non-adiabatic interaction matrix elements have positive values at their maximum magnitudes. Similar phase adjustments are implemented for the derived non-adiabatic interaction matrix elements from the valence-hole model (Fig.~\ref{fig:valence_hole_3Pig}c).

The $ab$ $initio$ calculations for the $^1\Sigma_u^+$ and $^3\Sigma_u^+$ states in Fig.~\ref{fig:sigma_valence_hole} are obtained using the CASSCF method. The CASSCF calculation is implemented with the ORCA quantum chemistry package~\cite{neese2020orca}. The full valence active space includes all of the MOs that arise from the 2s and 2p orbitals of the C atoms (cc-pVTZ basis).

\subsection{Global diabatization scheme}
\label{sec:global_scheme}

In the global diabatization schemes for the $^{1,3}\Pi_g$ and $^{1,3}\Sigma_u^+$ states obtained from the $ab$ $initio$ calculations, the valence-hole state interacts with each of the other diabats via the electrostatic interaction. In the $^1\Pi_g$ (Fig.~\ref{fig:valence_hole}) and $^{1,3}\Sigma_u^+$ (Fig.~\ref{fig:sigma_valence_hole}) manifolds, interactions among these other diabats are neglected, all of which are associated with electron configurations with four electrons in the valence-core. To model the $^3\Pi_g$ states (Fig.~\ref{fig:valence_hole_3Pig}), we include an additional interaction between the second and third diabats, which cross in the dashed boxed region of Fig.~\ref{fig:valence_hole_3Pig}a. This additional curve-crossing leads to the well-studied avoided-crossing between the $d^3\Pi_g$ and $e^3\Pi_g$ states~\cite{Ballik1963,phillips1949fox}. For a specified symmetry manifold, the valence-hole state is assumed to dissociate into the lowest possible 2s$^2$2p$^2$+2s2p$^3$ separated-atom limit allowed for that electronic-state symmetry. 


A four-parameter, Morse-$like$ potential~\cite{jia2012equivalence} is used to model each of the bound diabatic potentials in Figs.\,\ref{fig:valence_hole}\,-\,\ref{fig:sigma_valence_hole},
\begin{equation} \label{eq:bound}
T_e+D\left(1-\frac{e^{\beta R_e+h}}{e^{\beta R+h}}\right),
\end{equation}
where $D=E_{diss}-T_e$ is the dissociation energy for that potential curve relative to the minimum of the potential at $T_e$. The energies of various 2s$^2$2p$^2$+2s$^2$2p$^2$ fragment channels, $E_{diss}$, are fixed at the calculated values. The energies of the 2s$^2$2p$^2$($^3P$)+2s2p$^3$($^3D$) (Fig.~\ref{fig:valence_hole}) and 2s$^2$2p$^2$($^3P$)+2s2p$^3$($^5S$) (Fig.~\ref{fig:valence_hole_3Pig}a) channels are not provided by the $ab$ $initio$ calculations. These two excited-C channel energies are fixed at their respective experimental values in our global diabatic model. The empirical four-parameter function given by Eq.~\ref{eq:bound} is equivalent to both the Wei~\cite{hua1990four} and Tietz~\cite{tietz1963potential} potentials. It has the same functional form as the Morse potential when $h=0$. The use of these four-parameter potentials results in better agreement with the $ab$ $initio$ curves than the use of the three-parameter Morse potentials. The repulsive states are modeled by an exponential decay function, $\mathcal{A}_re^{-k_r R}$. 

To model the $ab$ $initio$ potentials for the entire calculated $R$-range with our diabatization scheme, the electronic part of the electrostatic interactions, $H_{ij}^{el}(R)$, between two specified diabatic states, $i$ and $j$, must vanish as $R\rightarrow \infty$. This ensures that the diabatic and adiabatic states converge to the same energy for each dissociation channel. For simplicity, we assume that all of the $H_{ij}^{el}(R)$ interaction terms decay exponentially with $R$ in the fits to the $ab$ $initio$ results, i.e., $H_{ij}^{el}(R)=\mathcal{H}_{ij}e^{-s_{ij} R}$. Within a specified symmetry manifold, the exponential decay rates, $s_{ij}$, for all pairs of electrostatic interactions, are assumed to be the same. Only the amplitudes, $\mathcal{H}_{ij}$, of these exponential-decay functions are individually determined from the fit. As with the diabatic fit model to the experimental level energies, to reproduce the $ab$ $initio$ calculations, the specific choice of $R$-dependent form of $H_{ij}^{el}$ is of importance secondary to the use of the correct curve-crossing patterns for the diabatization scheme. Even though the proposed exponential decay form of $H_{ij}^{el}(R)$ leads to unphysically large values of $H_{ij}^{el}$ at the short-$R$ region ($R\ll1\,\textup{\AA}$), this unphysical behavior in the electrostatic interaction terms does not significantly degrade the validity of our diabatization schemes. The avoided-crossing patterns are most strongly influenced by the shapes of the diabatic potentials and the magnitudes of $H_{ij}^{el}$ at the $R$-values of their intersections, all of which are in the $1.5\,\textup{\AA}\lesssim R\lesssim2\,\textup{\AA}$ range for the four symmetry species of C$_2$ in the energy region of interest (see Figs.\,\ref{fig:valence_hole}\,-\,\ref{fig:sigma_valence_hole}). Numerical details of various global valence-hole models are given in Section S4 of the supplementary text. 

\section{Diabatic two-state model for the $C^1\Pi_g$ state}
\label{sec:results}

The Morse potential parameters from the B-R model are listed in Table~\ref{tab:fit}. Despite its poorer numerical accuracy than the ``valence-hole'' two-state model (see Section S2 of the supplementary text for details), the B-R model captures the strongest diabatic interaction relevant to the $C^1\Pi_g$ state. According to our implementation of a B-R model, the curve-crossing between the two diabats occurs in the $v=6$ energy region of the d$_1$ state. The interaction matrix element between the two diabats, $H_e$ (Table~\ref{tab:fit}), is enormous. The magnitude of this interaction ($>$3000 cm$^{-1}$) is much larger than the vibrational energy spacings. The entire manifold of vibrational levels of both d$_1$ and d$_2$ diabats is strongly perturbed as a result of this interaction. This accounts for the significant deviation of the $C$-state energy level structure from that of a Morse-like potential at low $v$. 

\begin{table} 
\caption{Molecular parameters derived for the B-R model of the C$_2$ $C$ state. Based on the newly-measured C$_2$ bond dissociation energy~\cite{Borsovszky2021} and the $X$-state molecular constants~\cite{chen2015simultaneous}, the C($^3P$)+C($^3P$) limit is taken to be 51315 cm$^{-1}$ above the minimum of the $X$-state potential. All of the specified carbon atomic states belong to the 2s$^2$2p$^2$ electron configuration. Numbers in parentheses are $1\sigma$ uncertainties of the last digits.}
\begin{center}
\begin{tabular}{@{\hspace{8pt}} c @{\hspace{8pt}} | @{\hspace{6pt}} c @{\hspace{6pt}}  @{\hspace{6pt}} c   @{\hspace{6pt}}c  @{\hspace{6pt}} c  @{\hspace{6pt}}  @{\hspace{6pt}} c   @{\hspace{6pt}}c   @{\hspace{6pt}} c @{\hspace{6pt}} @{\hspace{6pt}} c   @{\hspace{6pt}}c  | @{\hspace{6pt}} c @{\hspace{6pt}} | @{\hspace{6pt}} c @{\hspace{6pt}}}

\hline
Diabatic state & $T_e$\,/\,cm$^{-1}$ & $\beta$\,/\,$\text{\AA}^{-1}$ & & $R_e\,/\,\text{\AA}$& Dissociation Limit\\
\hline
d$_1$ & 34580.6(31) & 2.9269(21) & &	1.2442(4) & C($^1D$)+C($^1D$) \\
d$_2$& 45371.6(714) & 3.2822(338)& & 1.5738(40)& C($^3P$)+C($^3P$) \\
\hline
$H_e$\,/\,cm$^{-1}$ & 3322.30(642) & && &\\
\hline
\end{tabular}
\end{center}
\label{tab:fit}
\end{table}

An alternative explanation of the unusual $C$-state level structure was proposed by Messerle and Krauss~\cite{Messerle1967}. They proposed that the $C^1\Pi_g$ state is perturbed by a $C'^1\Pi_g$ state, which has its minimum near the energy of the $C$-state $v=3$ level (Fig.~\ref{fig:PECs}b). In that assignment scheme, the $C$-state $v=n$ level ($n\geq3$) is locally (within an energy difference of $\sim$10 cm$^{-1}$) perturbed by the $C'$-state $v=n-3$ level. We claim that the Messerle and Krauss interpretations of the observed $^1\Pi_g$ energy level structure are incorrect, because they are based on unphysical potential curves and transition dipole moments (see discussions in Section S1 of the supplementary text). As is evident in Fig.~\ref{fig:PECs}b, instead of the expected repulsive, avoided-crossing pattern, the $C$- and $C'$-state PECs are treated as closely parallel with each other in the 42000$-$46000 cm$^{-1}$ region. Our $ab$ $initio$ calculation (Fig.~\ref{fig:valence_hole}) demonstrates that the second $adiabatic$ $^1\Pi_g$ state lies much higher in energy than this incorrectly-proposed $C'^1\Pi_g$ state, in good agreement with the B-R model (the a$_2$ state in Fig.~\ref{fig:PECs}a) and the $ab$ $initio$ results from Ref.~\citenum{schmidt2021spectroscopy}. The Messerle-Krauss band system ($C'^1\Pi_g-A^1\Pi_u$) has frequently been included as one of the electronic band systems of C$_2$, for example, in Ref.~\citenum{Huber1979} by Huber and Herzberg, as well as in more recent C$_2$ line-lists (Ref.~\citenum{furtenbacher2016experimental}). Based on the results of the B-R model and the $ab$ $initio$ calculation, we conclude that the $C'$ state proposed by Messerle and Krauss does not exist.

\section{Valence-hole diabatic states}
\label{sec:vh}

The equilibrium energy of the d$_1$ state is the lowest among the five diabats in our global diabatization scheme for the $^1\Pi_g$ states, and yet, the d$_1$ state dissociates into a qualitatively different fragment channel which lies at a much higher energy than the other diabats (Fig.~\ref{fig:valence_hole}). The binding energy of this d$_1$ state is $\sim$80000 cm$^{-1}$. This exceptional stability of the d$_1$ state relative to its dissociation products can be explained by the electron configuration associated with the d$_1$ state. In our diabatic model, the d$_1$ state accounts for $>$90$\%$ of the $C$-state character at the minimum of its potential. The dominant electron configuration at the $C$-state equilibrium is $2\sigma_g^22\sigma_u^11\pi_{u}^33\sigma_g^2$. This ``valence-hole'' configuration has a nominal bond order (BO) of three, consistent with the large binding energy of the d$_1$ state. In addition, the valence-hole configuration correlates with one 2p$\leftarrow$2s promoted fragment channel.

The ``valence-hole'' concept has been previously invoked by \citeauthor{lewis2005predissociation} in their analysis of the $b^1\Pi_u$ and $C^3\Pi_u$ states of N$_2$~\cite{lewis2005predissociation,lewis2008coupled}. \citeauthor{lewis2005predissociation} specify that the ``unusually shaped'' PECs for these two states arise from the diabatic interaction between a valence-hole configuration, $2\sigma_g^22\sigma_u^13\sigma_g^21\pi_{u}^41\pi_{g}^1$ (BO of 3), and $2\sigma_g^22\sigma_u^23\sigma_g^11\pi_{u}^31\pi_{g}^2$ (BO of 1). The authors deperturb the effect of the valence-hole induced curve-crossing on the lowest-energy $^{1,3}\Pi_u$ states, but do not implement a global diabatic deperturbation of the $^{1,3}\Pi_u$ electronic structure.

Compared to the other five C/N/O diatomic molecules, the valence-hole states in C$_2$, and their curve-crossings with other valence states occur at a much lower energy region. Rydberg-valence interactions, which affect the valence-hole state in these other C/N/O diatomics (such as N$_2$)~\cite{lewis2005predissociation,lewis2008coupled}, do not directly affect the valence-hole-induced curve-crossings in C$_2$, due to the uniquely low-lying nature of the valence-hole states of C$_2$. The effects of the valence-hole configurations on the global electronic structure are thus gloriously and uniquely sampled by the molecular constants of C$_2$. 

In our diabatic picture for the global electronic structure of the C$_2$ $^1\Pi_g$ states, the association of the d$_1$ state with the valence-hole configuration provides a chemically-intuitive explanation for $all$ $of$ $the$ $assumed$ $curve$-$crossings$ shown in Fig.~\ref{fig:valence_hole}. Based on the electron configuration analysis, the other four diabats in our model are composed of the $normal$ valence configurations with a quadruply-occupied valence-core. One electron in these configurations occupies one of the anti-bonding orbitals ($1\pi_{g}$ or $3\sigma_u$). As a result, these $normal$ valence states are either weakly bound or repulsive and dissociate into the lower-energy fragment channels that do not involve an 2p$\leftarrow$2s promotion. As $R$ increases, the deeply-bound, valence-hole state crosses all four of these other normal valence states. 
\bigbreak
As demonstrated in the following two sections, the valence-hole curve-crossing pattern for the $^1\Pi_g$ states (Fig.~\ref{fig:valence_hole}) can also be applied as a model for the electronic structure of the low-lying $^3\Pi_g$ and $^{1,3}\Sigma_u^+$ states of C$_2$. 

\subsection{Global diabatization scheme for the $^3\Pi_g$ states}
\label{sec:3pi_states}

\begin{figure}[t]
\center
\includegraphics[width=6.3 in]{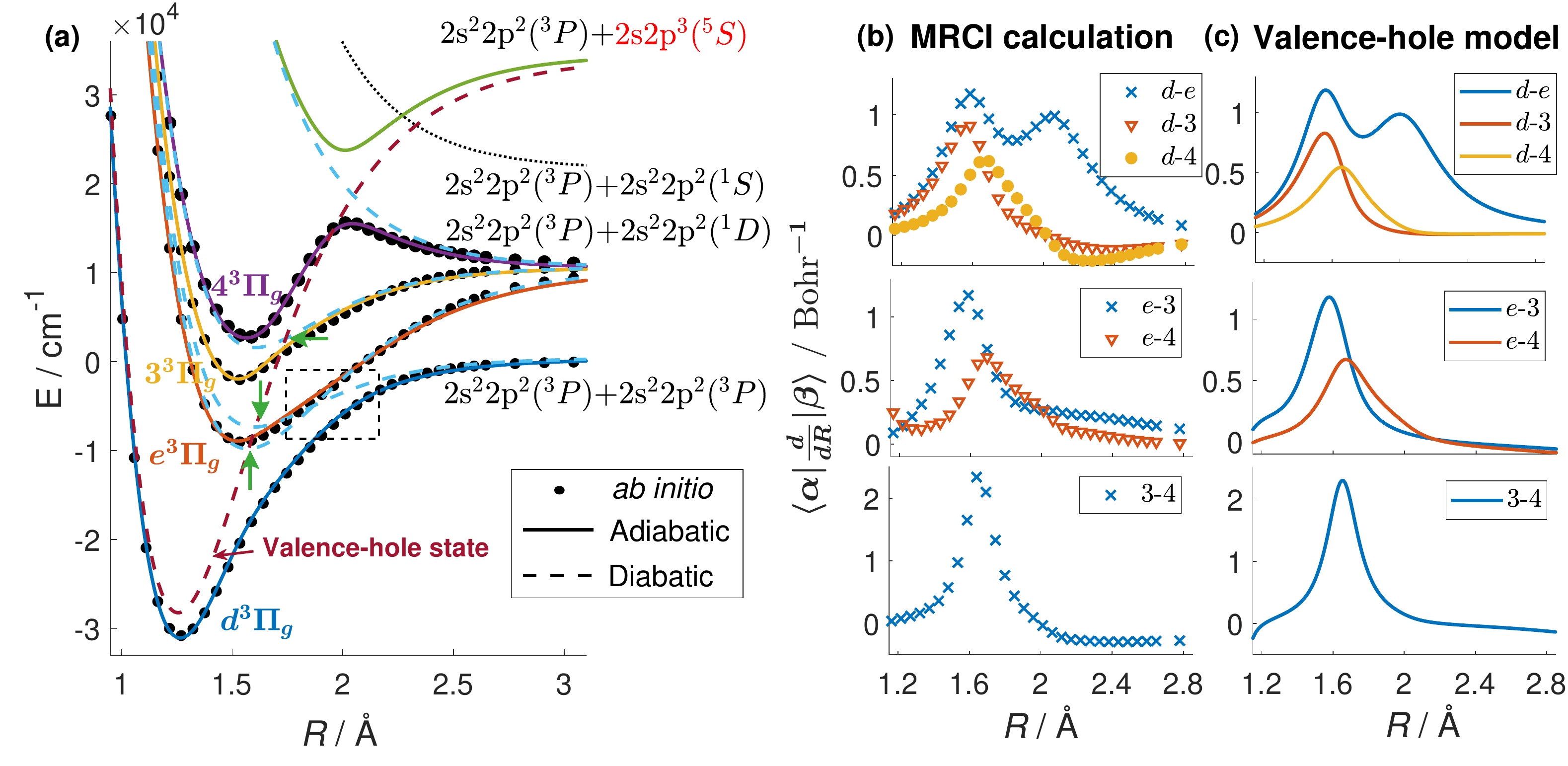}
\caption{The global diabatization scheme for the $^3\Pi_g$ states (see Section~\ref{sec:theory} for details). (a) The potential energy curves for the lowest $^3\Pi_g$ states from the $ab$ $initio$ calculation and our global diabatic interaction model. The $ab$ $initio$ potentials are calculated using the multi-reference configuration interaction (MRCI) method. Based on inspection of the potential curves from a CASSCF-level calculation (Fig.~S4b), the diabatic state that converges to the 2s$^2$2p$^2$($^3P$)+2s$^2$2p$^2$($^1S$) limit (illustrated qualitatively by the dotted line in Fig.~\ref{fig:valence_hole_3Pig}a) is repulsive, and interacts weakly with the valence-hole state. This interaction is neglected in our global diabatic model. (b) and (c) Illustration of the validity of the global diabatic deperturbation of the low-lying $^3\Pi_g$ states. The pairwise non-adiabatic interaction matrix elements for the four $^3\Pi_g$ states from the MRCI calculation are shown in panel (b). The corresponding $R$-dependent matrix elements derived from the valence-hole model are shown in panel (c).}
\label{fig:valence_hole_3Pig}
\end{figure}

The $ab$ $initio$ potentials of the lowest four $^3\Pi_g$ states (Fig.~\ref{fig:valence_hole_3Pig}a) are well reproduced by a similar diabatization scheme in which the deeply-bound valence-hole state crosses multiple weakly-bound/repulsive states before reaching a high-energy 2s$^2$2p$^2$($^3P$)+$2s2p^3$($^5S$) fragment limit. As a result of the proximity of curve-crossings in both energy- and $R$-values, the multiple-curve-crossing pattern is less obvious in the $^3\Pi_g$ manifold than in the $^1\Pi_g$ manifold. The $^3\Pi_g$ valence-hole state crosses each of the 2$^{\textup{nd}}$, 3$^{\textup{rd}}$, and 4$^{\textup{th}}$ $normal$ valence states (indicated by the three green arrows in Fig.~\ref{fig:valence_hole_3Pig}a) in about one half as large an energy- and $R$-region as its singlet counterpart (Fig.~\ref{fig:valence_hole}). To demonstrate the validity of our valence-hole diabatization scheme for the $^3\Pi_g$ states, we show, in Figs.~\ref{fig:valence_hole_3Pig}b and \ref{fig:valence_hole_3Pig}c, the pairwise non-adiabatic interaction matrix elements, $\langle\alpha| \frac{d}{dR}| \beta\rangle$, for the four adiabatic $^3\Pi_g$ states derived from the MRCI calculation and our valence-hole model. 

As is evident from Figs.~\ref{fig:valence_hole_3Pig}b and \ref{fig:valence_hole_3Pig}c, our diabatic model reproduces the strong $R$-dependence of all six non-adiabatic matrix elements from the MRCI calculation, including their maximum magnitudes and their $R$-values. According to the valence-hole model, the maximum value at $R\sim2$\,\textup{\AA} in the double-humped $\langle d| \frac{d}{dR}| e\rangle$ matrix element is due to the curve-crossing between the two non-valence-hole states in the boxed region of Fig.~\ref{fig:valence_hole_3Pig}a. With the exception of this feature, all of the other peaks in the non-adiabatic matrix elements are caused by curve-crossings with a single valence-hole state. These curve-crossings lead to strong $R$-dependence in the valence-hole character on each of the consecutively higher-energy $^3\Pi_g$ states. The presence of the valence-hole state causes prolific, strongly $R$-dependent, non-adiabatic interactions among the low-lying $^3\Pi_g$ states.

The manifold of $^3\Pi_g$ electronic states includes the well-characterized $d^3\Pi_g$ and $e^3\Pi_g$ states, and the recently observed $3^3\Pi_g$ and $4^3\Pi_g$ states~\cite{krechkivska2015resonance,krechkivska2017first}. With the large amount of spectroscopic data across four different electronic states, we are able to construct a complete $global$ $spectroscopic$ fit model for the $^3\Pi_g$ states. This model reproduces all of the observed level energies with an average fit residual of $<$0.5 cm$^{-1}$. This fit model for the vibration-rotation levels of the four $^3\Pi_g$ states will be reported in a future paper. 

\subsection{Global diabatization scheme for the $^{1}\Sigma_u^+$ and $^{3}\Sigma_u^+$ states}
\label{sec:sigma_states}

\begin{figure}
\center
\includegraphics[width=6 in]{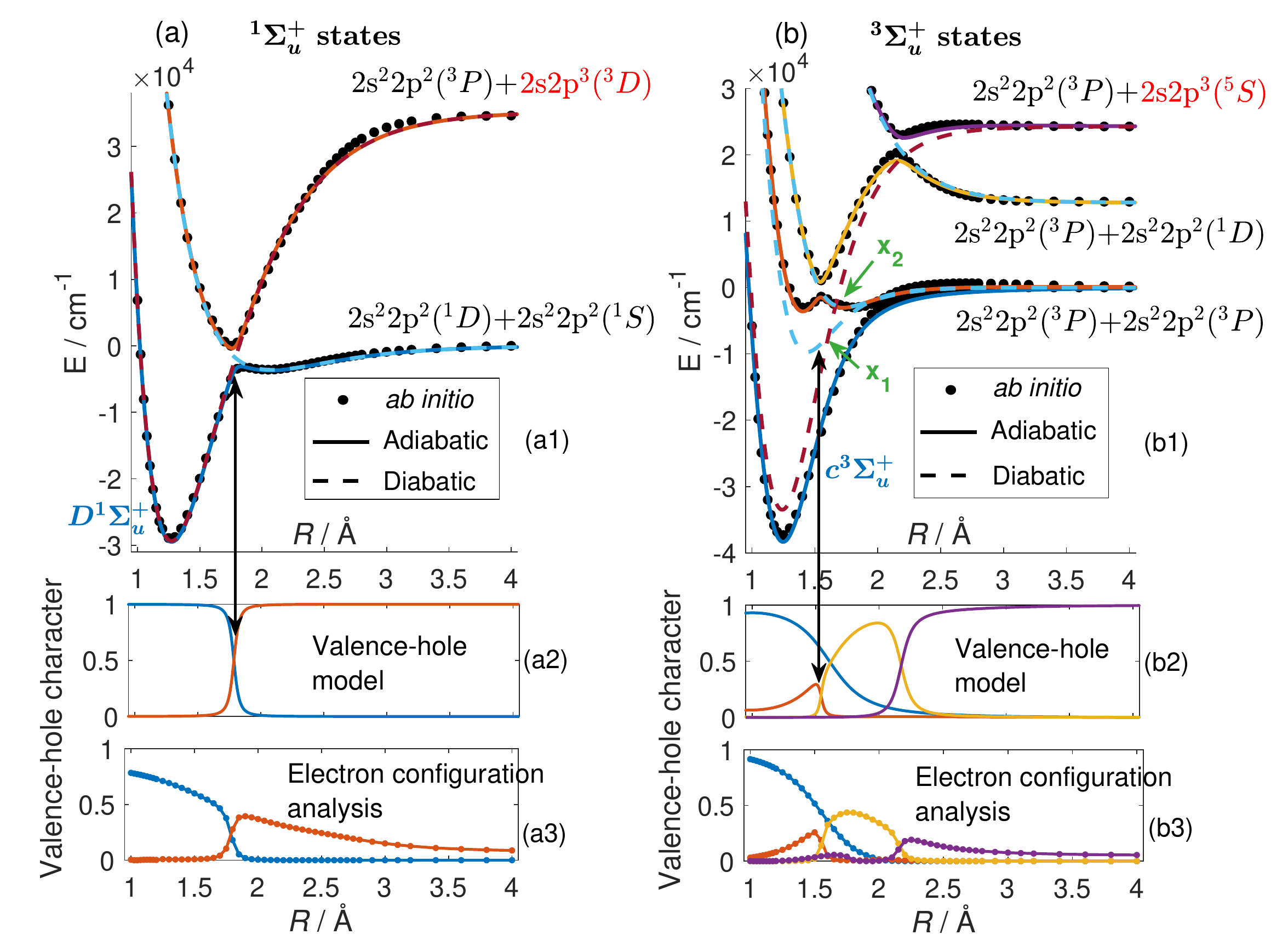}
\caption{The $ab$ $initio$ results, calculated with the CASSCF method, for the low-lying $^{1}\Sigma_u^+$ and $^{3}\Sigma_u^+$ states (see Section~\ref{sec:theory} for details). Our global diabatization schemes, based on the valence-hole concept, for the electronic structure of the two symmetry species are shown in panels (a1) and (b1). The valence-hole character from the valence-hole model and the electron configuration analysis based on the CASSCF calculation are shown, respectively, in the lower two panels of each column, column (a) for the $^{1}\Sigma_u^+$ states and column (b) for the $^{3}\Sigma_u^+$ states.}
\label{fig:sigma_valence_hole}
\end{figure}

The global diabatization schemes for the low-lying $^{1,3}\Sigma_u^+$ potentials are shown in the two upper panels of Fig.~\ref{fig:sigma_valence_hole}. At $R_e$, the $D^{1}\Sigma_u^+$ and $c^{3}\Sigma_u^+$ states are both dominated by the $2\sigma_g^22\sigma_u^11\pi_{u}^43\sigma_g^1$ valence-hole configuration. Our diabatic model predicts that the valence-hole character is transferred between neighboring adiabatic states at each successsive diabatic curve-crossing. The point of transfer of the valence-hole character between the lowest two $^1\Sigma_u^+$ states is indicated on Fig.~\ref{fig:sigma_valence_hole}a by a double-headed arrow. The curve-crossing pattern and the resulting diabatic interactions are more complicated among the $^3\Sigma_u^+$ states. As a result of the proximity in both energy- and $R$-values of the first two valence-hole curve-crossings (indicated by x$_1$ and x$_2$ in Fig.~\ref{fig:sigma_valence_hole}b1), the second adiabatic $^3\Sigma_u^+$ state loses its valence-hole character to the third state (as shown by the double-headed arrow in Fig.~\ref{fig:sigma_valence_hole}b) even before the first curve-crossing (x$_1$).

This pattern of transfer of valence-hole character across adiabatic states implied by our valence-hole model is confirmed by the electron configuration analysis based on the CASSCF calculation. In Figs.~\ref{fig:sigma_valence_hole}a3 and \ref{fig:sigma_valence_hole}b3, the contribution (i.e. amplitude squared) of the valence-hole configuration ($2\sigma_g^22\sigma_u^11\pi_{u}^43\sigma_g^1$) in the total CASSCF wavefunction is shown as a function of $R$ for the low-lying $^{1,3}\Sigma_u^+$ states. As is evident from the electron configuration analysis, the valence-hole configuration jumps between neighboring adiabatic states at the same internuclear distance regions predicted by the valence-hole model. Note that the contribution from the $2\sigma_g^22\sigma_u^11\pi_{u}^43\sigma_g^1$ configuration in the total CASSCF wavefunction decreases as $R$ increases. This dilution of the contribution from the $2\sigma_g^22\sigma_u^11\pi_{u}^43\sigma_g^1$ configuration at large $R$ is due to the increasing number of near-degenerate valence-hole states (i.e. with either $2\sigma_g^22\sigma_u^1$ or $2\sigma_g^12\sigma_u^2$ valence-core). For example, as $R\rightarrow\infty$, electron configurations such as $2\sigma_g^12\sigma_u^21\pi_{u}^33\sigma_g^11\pi_{g}^1$ and $2\sigma_g^22\sigma_u^11\pi_{u}^23\sigma_g^11\pi_{g}^2$ are energetically indistinguishable from $2\sigma_g^22\sigma_u^11\pi_{u}^43\sigma_g^1$.  Despite the crudeness of the LCAO-MO theory in describing the wavefunction at large $R$, the electron configuration analysis from the CASSCF calculation supports the valence-hole-induced curve-crossing patterns in our diabatic interaction model of the low-lying $^{1,3}\Sigma_u^+$ states.

\begin{figure}
\center
\includegraphics[width=5.5 in]{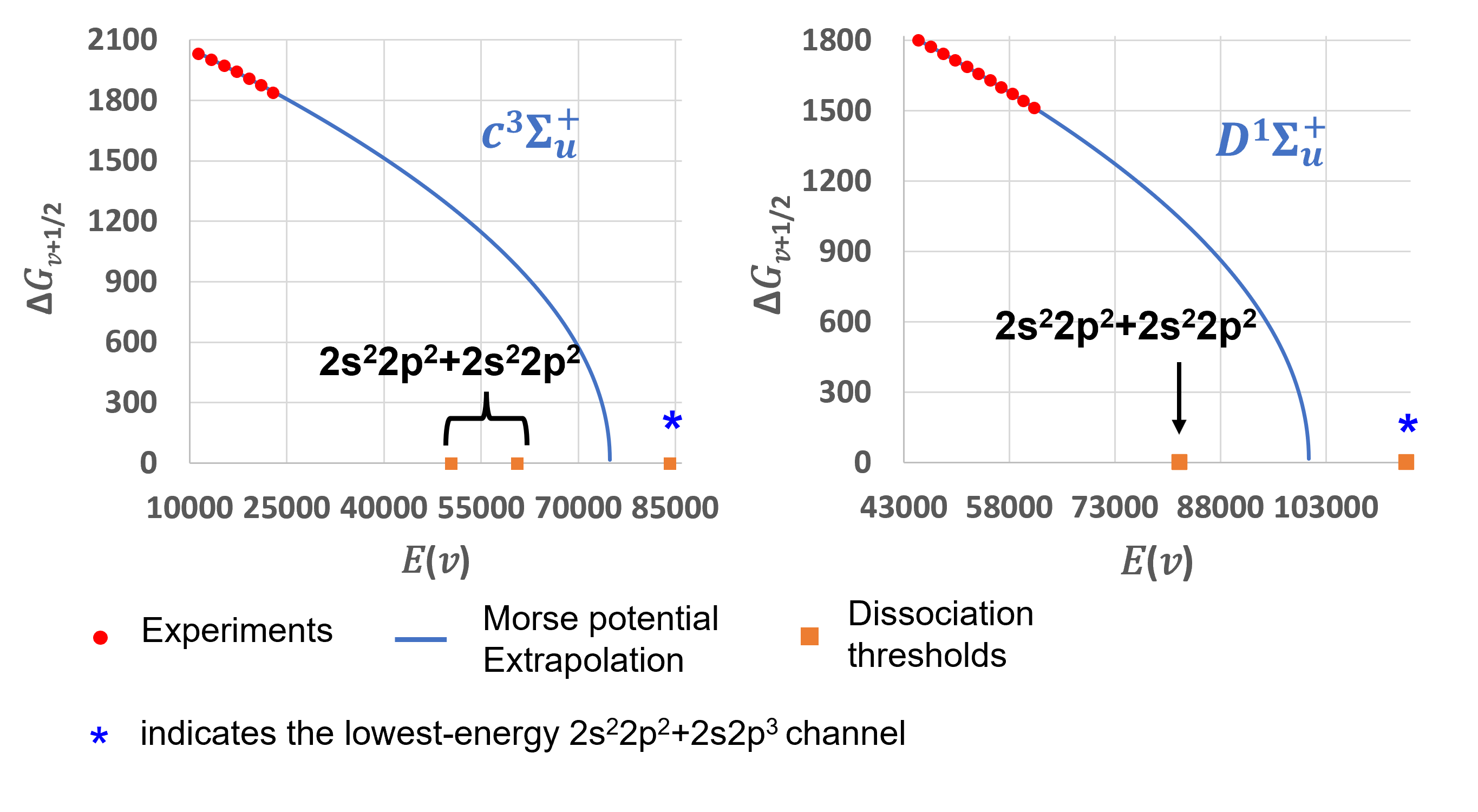}
\caption{The observed vibrational energy spacings, $\Delta G_{v+1/2}$, as a function of the vibrational energy levels, $E(v)$, for the $c^{3}\Sigma_u^+$ and $D^{1}\Sigma_u^+$ states.}
\label{fig:sigma_dissociation_trend}
\end{figure}

Due to the relatively high energy of the lowest-energy curve-crossing ($\sim$25000 cm$^{-1}$ above the minimum of the valence-hole diabatic potential) in the $^{1,3}\Sigma_u^+$ symmetry manifolds, the effects that result from the curve crossings by the valence-hole potential curves are not obvious in the observed molecular constants of the C$_2$ $D^{1}\Sigma_u^+$ and $c^{3}\Sigma_u^+$ states~\cite{blunt1995revised,krechkivska2018higher,joester2007d,chan2013laser,nakajima2013spectroscopic,nakajima2014excitation}. The trends in the observed vibrational energy spacings for these two electronic states, $\Delta G_{v+1/2}$, suggest that, in the diabatic picture, both the $D$ and $c$ states “want” to dissociate into a high-energy separated-atom limit. The $\Delta G_{v+1/2}$ values for the two states are shown as a function of the energy of the vibrational level energies, $E(v)$, in Fig.~\ref{fig:sigma_dissociation_trend}. The blue curves in the two panels of Fig.~\ref{fig:sigma_dissociation_trend} are obtained by fitting the available experimental values to a Morse potential form. While a Morse-type extrapolation from the $\Delta G_{v+1/2}$ values at low $v$ is not expected to yield a quantitatively accurate dissociation energy, it appears that the $\Delta G_{v+1/2}$ values of both $D$ and $c$ states do not decrease rapidly enough for either state to converge to one of the 2s$^2$2p$^2$+2s$^2$2p$^2$ separated-atom configurations, in agreement with our assumption about the high dissociation energy for the diabatic $^{1,3}\Sigma_u^+$ valence-hole states.

\section{Valence-hole $vs$. quadruple-bonding in C$_2$}
\label{sec:quadruple}

The quadruple bond for the C$_2$ $X^1\Sigma_g^+$ state, proposed by \citeauthor{shaik2012quadruple}~\cite{shaik2012quadruple,shaik2013one,shaik2016quadruple}, has been rationalized by assuming a high ``intrinsic bonding energy'' (106000 cm$^{-1}$) for the $X$ state at $R_e$~\cite{shaik2012quadruple,shaik2013one}. This $quadruply$-bonded $X$ state correlates with the $doubly$-promoted, 2s2p$^3(^5S$)+2s2p$^3(^5S$) limit. While it may appear that the quadruply-bonded $X$ state shares bonding properties similar to those of our valence-hole states, the presence of the valence-hole states has a more profound global impact on the electronic structure of C$_2$ than the proposed quadruple-bonding, even though the quadruply-bonded artifact is mostly manifest in the electronic ground state. Instead of providing clarifications to the electronic structure of C$_2$, the quadruple-bonding model $creates$ additional obstacles to our intuitive understanding of the electronic structure theory. For example, this valence-bond-theory model fails to provide a simple picture that aids intuitive rationalization of the differences between the molecular constants of the C$_2$ $X^1\Sigma_g^+$ state and those of the C$_2$ $c^{3}\Sigma_u^+$ state and the acetylene $X^1\Sigma_g^+$ state. These differences in the values of the molecular constants can be qualitatively understood using simple MO-theory concepts, such as the valence-hole configuration. 


\subsection{Global $vs$. localized effect}  
\label{sec:not_global}

In contrast with the valence-hole states that create systematic havoc in the global electronic structure of C$_2$, the quadruply-bonded character of the $X$ state is the ``victim'' rather than the ``instigator'' of the ``chaos'', which is rampant in C$_2$. The dominant electron configuration of the $X$ state at $R_e$, $2\sigma_g^22\sigma_u^21\pi_{u}^4$, correlates with a triply-bonded structure, due to the weakly anti-bonding nature of the $2\sigma_u$ MO, the weakness of which is the result of significant sp-hybridization in C$_2$ at short internuclear distances~\cite{shaik2012quadruple,shaik2013one,su2011bonding,shaik2016quadruple}. In MO theory~\cite{shaik2012quadruple,shaik2013one}, the proposed fourth bond for the C$_2$ $X$ state, which has a bond strength similar to that of a hydrogen bond, results from a weak interaction of the triple-bond structure with a ``double-hole'' configuration, $2\sigma_g^21\pi_{u}^43\sigma_g^2$. This double-hole configuration correlates with a true quadruply-bonded structure. No diabatic curve-crossing occurs for this configuration interaction, because this double-hole state lies much higher in energy than $2\sigma_g^22\sigma_u^21\pi_{u}^4$. 

The proposed quadruply-bonded structure of the $X$ state at $R_e$ (i.e. a strong triple-bond + a weak fourth bond) has minimal impact on the global electronic structure of C$_2$, because the quadruple-bonding effect is localized near $R_e$. As is evident from the valence-bond theory calculation for the $X$ state~\cite{su2011bonding}, the diabatic curve for the valence-bond structure with the triple-bond character at $R_e$ has a well-depth of only 50000 cm$^{-1}$. The triple-bond component of the quadruply-bonded structure morphs into a double-bond at larger $R$, because the sp-hybridization demanded by the valence-bond picture decays as $R$ increases. The quadruply-bonded structure does not dissociate into a 2s2p$^3$+2s2p$^3$ double-valence-hole limit, because the quadruply-bonded character disappears long before the molecule reaches even the lowest-energy 2s$^2$2p$^2$($^3P$)+2s$^2$2p$^2$($^3P$) fragment limit. 

Note that, in Fig. 1 of Ref.~\citenum{shaik2016quadruple} by \citeauthor{shaik2016quadruple}, it is shown that the dominant valence-bond character of the C$_2$ $X$ state changes from a quadruply-bonded structure at $R_e$ to a linear combination of various doubly-bonded structures at larger $R$ ($\gtrsim1.85\,\textup{\AA}$). This $single$ cross-over of dominant bonding character is taken as a ``clear justification" for the exceptionally high dissociation energy (106000 cm$^{-1}$) of the quadruply-bonded structure at $R_e$~\cite{shaik2016quadruple}. We disagree with this argument. As discussed earlier, the assumed high dissociation energy for the proposed quadruply-bonded structure of the C$_2$ $X$ state is in clear contradiction with the considerably lower binding-energy (50000 cm$^{-1}$) of its triple-bond component at $R_e$, as derived in Ref.~\citenum{su2011bonding}. In addition, the diabatic dissociation limit of the quadruply-bonded structure cannot be reliably inferred from the $R$-dependent quadruple-bonding character on a $single$ adiabatic state, which lies $>$100000 cm$^{-1}$ lower in energy than the assumed 2s2p$^3(^5S$)+2s2p$^3(^5S$) fragment limit for such a quadruply-bonded structure.

The cross-over of bonding character at $R\sim1.85\,\textup{\AA}$ for the $X$ state is an expected result, given the presence of the well-documented avoided-crossing between the $X^1\Sigma_g^+$ and $B'^1\Sigma_g^+$ states at that $R$-region~\cite{abrams2004full,varandas2008extrapolation}. As discussed below, the diabatic curve-crossing that leads to this exchange of dominant bonding character can be qualitatively explained by simple MO-theory arguments, without invoking a quadruple-bonding proposal for the $X$ state. 

The lowest-energy, 2s$^2$2p$^2$($^3P$)+2s$^2$2p$^2$($^3P$) fragment limit of C$_2$ correlates with two $^1\Sigma_g^+$ states~\cite{Ballik1963,herzberg1950molecular}. The higher-energy $B'$ state ($T_e=12082\,\textup{cm}^{-1}$) has a larger $R_e$ (1.377\,\textup{\AA}) and smaller vibrational frequency (1407\,cm$^{-1}$) than the $X$ state (1.243\,\textup{\AA}, 1855\,cm$^{-1}$)~\cite{douay1988discovery,chen2015simultaneous}. This ensures that the diabatic $X$- and $B'$-state PECs cross at some large $R$ value ($\sim$1.6\,\textup{\AA}), according to a two-state diabatic model for the $ab$ $initio$ PECs of the two lowest $^1\Sigma_g^+$ states, calculated with the MRCI method~\cite{varandas2008extrapolation}. This curve-crossing occurs because of the initial shapes and relative energies of the two diabatic potentials, in combination with the electronic-state symmetry requirement at bond dissociation. The dominant electron configuration of the $B'$ state at $R_e$ ($2\sigma_g^22\sigma_u^21\pi_{u}^23\sigma_g^2$) derives from a double $3\sigma_g\leftarrow 1\pi_u$ electron promotion from the dominant configuration of the $X$ state at $R_e$~\cite{Chabalowski1983}. The difference in the shapes and energies between the $X$- and $B'$-state PECs can be qualitatively understood, because the $1\pi_u$ MO is more strongly-bonded than the $3\sigma_g$ MO. 

\subsection{The quadruple-bonding picture and chemical intuitions}
\label{sec:no_insights}

The quadruple-bonding model of the C$_2$ $X$ state is proposed at the expense of chemical intuition, which, ironically, is one of the potential gains from using the valence-bond theory over the traditional LCAO-MO approach. Despite its proposed $quadruply$-bonded character, the $R_e$ of the C$_2$ $X$ state is longer than that of the CC $triple$-bond of the electronic ground state of acetylene ($X^1\Sigma_g^+$). In addition, the force constant of the C$_2$ $X$ state is smaller than that of the CC triple-bond of acetylene. In order to reconcile these two discordant experimental facts with their quadruple-bonding model, \citeauthor{shaik2016quadruple}~\cite{shaik2016quadruple} argue that the $\sigma$-bond components of the quadruply-bonded C$_2$ $X$ state ``prefer" a larger $R$ value than the CC $\sigma$-bond in acetylene, as a result of Pauli repulsion between the two $\sigma$-bonds in the C$_2$ $X$ state. Using a similar argument, \citeauthor{shaik2016quadruple} rationalize the difference in the $R_e$ and force constant between the C$_2$ $X^1\Sigma_g^+$ and $c^3\Sigma_u^+$ states~\cite{shaik2012quadruple,shaik2013one,shaik2016quadruple}. According to their valence-bond theory argument, the two $\sigma$-bonds in the quadruply-bonded $X$ state evolve into a combination of a single $\sigma$-bond and a non-interacting triplet electron pair at the $R_e$ of the $c$ state. \citeauthor{shaik2016quadruple} claim that, due to reduced spatial congestion for this $\sigma$-bond\,+\,triplet-pair configuration compared with a double $\sigma$-bond, the $R_e$ and force constant of the $triply$-bonded $c$ state should be expected to be shorter and larger, respectively, than the corresponding values for the proposed $quadruply$-bonded $X$ state.

We suggest that these counter-intuitive valence-bond theory arguments of Refs. \citenum{shaik2012quadruple,shaik2013one,shaik2016quadruple} are unnecessary and fail the Occam's razor test, in particular those that rationalize the difference in the $R_e$ and force constant between the C$_2$ $X$ and $c$ states. This difference can be satisfactorily accounted for using simple MO-theory concepts, such as the valence-hole configuration. As discussed in Section \ref{sec:sigma_states}, the $c$ state is dominated by a valence-hole configuration at $R_e$, $2\sigma_g^22\sigma_u^11\pi_{u}^43\sigma_g^1$. The presence of this strongly-bound valence-hole configuration has a profound impact on the global electronic structure landscape of the $^3\Sigma_u^+$ states, as demonstrated in Fig.~\ref{fig:sigma_valence_hole}b by our multi-state diabatic model. The specialness of this valence-hole configuration, which derives from an bonding$\leftarrow$anti-bonding, $3\sigma_g\leftarrow2\sigma_u$ promotion from the dominant configuration of the $X$ state at $R_e$ ($2\sigma_g^22\sigma_u^21\pi_{u}^4$), is the simple qualitative reason for the difference between the $R_e$ and force constant of the $X$ and $c$ states, as well as their relative energies.

\bigbreak
In marked contrast to the diabatic valence-hole states, the quadruply-bonded character for the $X$ state at the $R_e$ region does not lead to spectroscopically-observable features in the electronic spectra of C$_2$, with the possible exception of the low-energy curve-crossing between the diabatic $X$ and $B'$ states, which is expected given the multi-configurational nature of C$_2$. Without any spectroscopic signatures, the quadruple-bonding model of the C$_2$ $X$ state has no physical relevance, and muddies the already complicated chemical-bonding picture of the molecule.

\section{Implications and outlook}
\label{sec:conclusions}

As the result of our effort to improve the numerical accuracy of the two-state model for the C$_2$ $C^1\Pi_g$ state, the special role of the ``valence-hole'' configuration in the global electronic structure of C$_2$ is revealed. The diabatic valence-hole curve-crossings in the $^{1,3}\Pi_g$ and $^{1,3}\Sigma_u^+$ symmetry manifolds of C$_2$ are analogous to the well-studied ionic (A$^+$B$^-$)/covalent (AB) crossings in strongly ionic species~\cite{herzberg1950molecular, zewail2000femtochemistry}. In both cases, the $global$ electronic structure is systematically disrupted by a single diabatic electronic state (ionic or valence-hole), which not only has an intrinsically larger binding energy than the other valence states in the same energy region, but also dissociates into a distinctly different, higher-energy channel (A$^+$+B$^-$ or 2s$^2$2p$^2$+2s2p$^3$). 

The valence-hole concept is a new arrow in the quiver of intuitive electronic structure models. Just as with the ion-pair states, the diabatic picture is the key to uncovering the disruptive impact of the valence-hole state on the global electronic structure landscape of C$_2$. The diabatic model, combined with the traditional MO concepts of bond order and correlation diagrams, disentangles the complex multi-configurational interactions in C$_2$. 

Our model, which explicitly treats the bound-continuum interaction, provides a compact and physical representation of unimolecular processes (e.g. predissociation) of the highly-excited states of C$_2$, as well as for the mechanisms of elementary C+C reactions and radiative association. The electronic- and vibrational-state-dependent predissociation rates of $^3\Pi_g$ vibration-rotation levels derived from our valence-hole model are consistent with the experimental observations~\cite{welsh2017pi,krechkivska2017first}. These calculated rates have been used in the modeling of the lifetimes for cometary C$_2$~\cite{Borsovszky2021}, which explains the astronomical observations. The valence-hole-induced vibronic mixing will also have significant impact on the state-specific collisional dynamics of the C$_2$ molecule~\cite{gelbart1973intramolecular}. Based on our demonstration of the importance of valence-hole configurations in C$_2$, we propose recognition and re-analysis of diabatic interactions in the other five C/N/O diatomic molecules, for which the valence-hole configurations will have a similar $global$ impact on their electronic structure, albeit in higher energy regions.

\begin{acknowledgement}

We thank Brenton Lewis (Australian National University) for valuable discussions during the project. This work was supported by the U.S. Department of Energy, Office of Science, Chemical Sciences, Geosciences, and Biosciences Division of the Basic Energy Sciences Office, under Award Number DE-FG0287ER13671; the U.S. National Science Foundation (CHE-1900358); Australian Research Council (DP190103151), with the assistance of resources from the National Computational Infrastructure, which is supported by the Australian Government (Project zu57, Centre of Excellence in Exciton Science CE170100026). RWF thanks the U.S. National Science Foundation (CHE-1800410) for support of his part of this collaboration. The manuscript preparation was carried out at the Lawrence Livermore National Laboratory under the auspices of the U.S. Department of Energy by Lawrence Livermore National Laboratory under Contract DE-AC52-07NA27344.

\end{acknowledgement}

\begin{suppinfo}
Supplemental Information: Additional discussions on the non-existence of the Messerle and Krauss $C'^1\Pi_g$ state (Section S1 of the supplementary text); Details of the valence-hole two-state fit model of the C$_2$ $C^1\Pi_g$ state (Section S2); CASSCF calculation for the C$_2$ $^1\Pi_g$ and $^3\Pi_g$ states (Section S3); Numerical details of various global diabatic models (Section S4). Results of the $ab$ $initio$ calculation are included in a separate Excel document.

\end{suppinfo}

\bibliography{C2_JPCA}

\providecommand{\latin}[1]{#1}
\makeatletter
\providecommand{\doi}
  {\begingroup\let\do\@makeother\dospecials
  \catcode`\{=1 \catcode`\}=2 \doi@aux}
\providecommand{\doi@aux}[1]{\endgroup\texttt{#1}}
\makeatother
\providecommand*\mcitethebibliography{\thebibliography}
\csname @ifundefined\endcsname{endmcitethebibliography}
  {\let\endmcitethebibliography\endthebibliography}{}
\begin{mcitethebibliography}{59}
\providecommand*\natexlab[1]{#1}
\providecommand*\mciteSetBstSublistMode[1]{}
\providecommand*\mciteSetBstMaxWidthForm[2]{}
\providecommand*\mciteBstWouldAddEndPuncttrue
  {\def\EndOfBibitem{\unskip.}}
\providecommand*\mciteBstWouldAddEndPunctfalse
  {\let\EndOfBibitem\relax}
\providecommand*\mciteSetBstMidEndSepPunct[3]{}
\providecommand*\mciteSetBstSublistLabelBeginEnd[3]{}
\providecommand*\EndOfBibitem{}
\mciteSetBstSublistMode{f}
\mciteSetBstMaxWidthForm{subitem}{(\alph{mcitesubitemcount})}
\mciteSetBstSublistLabelBeginEnd
  {\mcitemaxwidthsubitemform\space}
  {\relax}
  {\relax}

\bibitem[Gulania \latin{et~al.}(2019)Gulania, Jagau, and
  Krylov]{gulania2019eom}
Gulania,~S.; Jagau,~T.-C.; Krylov,~A.~I. EOM-CC guide to Fock-space travel: The
  C$_2$ edition. \emph{Faraday Discussions} \textbf{2019}, \emph{217},
  514--532\relax
\mciteBstWouldAddEndPuncttrue
\mciteSetBstMidEndSepPunct{\mcitedefaultmidpunct}
{\mcitedefaultendpunct}{\mcitedefaultseppunct}\relax
\EndOfBibitem
\bibitem[Hirsch \latin{et~al.}(1980)Hirsch, Bruna, Buenker, and
  Peyerimhoff]{hirsch1980non}
Hirsch,~G.; Bruna,~P.~J.; Buenker,~R.~J.; Peyerimhoff,~S.~D. Non-adiabatic
  coupling matrix elements $\langle\Psi^\alpha| \partial/\partial Q|
  \Psi^\beta\rangle$ for large CI wavefunctions. \emph{Chemical Physics}
  \textbf{1980}, \emph{45}, 335--347\relax
\mciteBstWouldAddEndPuncttrue
\mciteSetBstMidEndSepPunct{\mcitedefaultmidpunct}
{\mcitedefaultendpunct}{\mcitedefaultseppunct}\relax
\EndOfBibitem
\bibitem[Chabalowski \latin{et~al.}(1981)Chabalowski, Buenker, and
  Peyerimhoff]{Chabalowski1981}
Chabalowski,~C.~F.; Buenker,~R.~J.; Peyerimhoff,~S.~D. Theoretical study of the
  electronic transition moments for the $d^3\Pi_g-a ^3\Pi_u$ (Swan) and
  $e^3\Pi_g-a^3\Pi_u$ (Fox-Herzberg) bands in C$_2$. \emph{Chemical Physics
  Letters} \textbf{1981}, \emph{83}, 441--448\relax
\mciteBstWouldAddEndPuncttrue
\mciteSetBstMidEndSepPunct{\mcitedefaultmidpunct}
{\mcitedefaultendpunct}{\mcitedefaultseppunct}\relax
\EndOfBibitem
\bibitem[Chabalowski \latin{et~al.}(1983)Chabalowski, Peyerimhoff, and
  Buenker]{Chabalowski1983}
Chabalowski,~C.~F.; Peyerimhoff,~S.~D.; Buenker,~R.~J. The Ballik-Ramsay,
  Mulliken, Deslandres-d'Azambuja and Phillips systems in C$_2$: A theoretical
  study of their electronic transition moments. \emph{Chemical Physics}
  \textbf{1983}, \emph{81}, 57--72\relax
\mciteBstWouldAddEndPuncttrue
\mciteSetBstMidEndSepPunct{\mcitedefaultmidpunct}
{\mcitedefaultendpunct}{\mcitedefaultseppunct}\relax
\EndOfBibitem
\bibitem[Shaik \latin{et~al.}(2012)Shaik, Danovich, Wu, Su, Rzepa, and
  Hiberty]{shaik2012quadruple}
Shaik,~S.; Danovich,~D.; Wu,~W.; Su,~P.; Rzepa,~H.~S.; Hiberty,~P.~C. Quadruple
  bonding in C$_2$ and analogous eight-valence electron species. \emph{Nature
  Chemistry} \textbf{2012}, \emph{4}, 195--200\relax
\mciteBstWouldAddEndPuncttrue
\mciteSetBstMidEndSepPunct{\mcitedefaultmidpunct}
{\mcitedefaultendpunct}{\mcitedefaultseppunct}\relax
\EndOfBibitem
\bibitem[Shaik \latin{et~al.}(2013)Shaik, Rzepa, and Hoffmann]{shaik2013one}
Shaik,~S.; Rzepa,~H.~S.; Hoffmann,~R. One molecule, two atoms, three views,
  four bonds? \emph{Angewandte Chemie International Edition} \textbf{2013},
  \emph{52}, 3020--3033\relax
\mciteBstWouldAddEndPuncttrue
\mciteSetBstMidEndSepPunct{\mcitedefaultmidpunct}
{\mcitedefaultendpunct}{\mcitedefaultseppunct}\relax
\EndOfBibitem
\bibitem[Frenking and Hermann(2013)Frenking, and Hermann]{frenking2013critical}
Frenking,~G.; Hermann,~M. Critical comments on “One molecule, two atoms,
  three views, four bonds?”. \emph{Angewandte Chemie} \textbf{2013},
  \emph{125}, 6036--6039\relax
\mciteBstWouldAddEndPuncttrue
\mciteSetBstMidEndSepPunct{\mcitedefaultmidpunct}
{\mcitedefaultendpunct}{\mcitedefaultseppunct}\relax
\EndOfBibitem
\bibitem[Danovich \latin{et~al.}(2013)Danovich, Shaik, Rzepa, and
  Hoffmann]{danovich2013response}
Danovich,~D.; Shaik,~S.; Rzepa,~H.~S.; Hoffmann,~R. A response to the critical
  comments on “One molecule, two atoms, three views, four bonds?”.
  \emph{Angewandte Chemie International Edition} \textbf{2013}, \emph{52},
  5926--5928\relax
\mciteBstWouldAddEndPuncttrue
\mciteSetBstMidEndSepPunct{\mcitedefaultmidpunct}
{\mcitedefaultendpunct}{\mcitedefaultseppunct}\relax
\EndOfBibitem
\bibitem[Laws \latin{et~al.}(2019)Laws, Gibson, Lewis, and
  Field]{laws2019dicarbon}
Laws,~B.; Gibson,~S.; Lewis,~B.; Field,~R.~W. The dicarbon bonding puzzle
  viewed with photoelectron imaging. \emph{Nature Communications}
  \textbf{2019}, \emph{10}, 1--8\relax
\mciteBstWouldAddEndPuncttrue
\mciteSetBstMidEndSepPunct{\mcitedefaultmidpunct}
{\mcitedefaultendpunct}{\mcitedefaultseppunct}\relax
\EndOfBibitem
\bibitem[Deslandres and d'Azambuja(1905)Deslandres, and d'Azambuja]{DA1905}
Deslandres,~H.; d'Azambuja,~L. Variations des spectres de bandes du carbone
  avec la pression, et nouveaux spectres de bandes du carbone. \emph{Comptes
  Rendus} \textbf{1905}, \emph{140}, 917\relax
\mciteBstWouldAddEndPuncttrue
\mciteSetBstMidEndSepPunct{\mcitedefaultmidpunct}
{\mcitedefaultendpunct}{\mcitedefaultseppunct}\relax
\EndOfBibitem
\bibitem[Dieke and Lochte-Holtgreven(1930)Dieke, and
  Lochte-Holtgreven]{Dieke1930}
Dieke,~G.; Lochte-Holtgreven,~W. {\"U}ber einige Banden des
  Kohlenstoffimolek{\"u}ls. \emph{Zeitschrift f{\"u}r Physik} \textbf{1930},
  \emph{62}, 767--794\relax
\mciteBstWouldAddEndPuncttrue
\mciteSetBstMidEndSepPunct{\mcitedefaultmidpunct}
{\mcitedefaultendpunct}{\mcitedefaultseppunct}\relax
\EndOfBibitem
\bibitem[Herzberg and Sutton(1940)Herzberg, and Sutton]{Herzberg1940}
Herzberg,~G.; Sutton,~R. Tail bands of the Deslandres-d'Azambuja system of the
  C$_2$ molecule. \emph{Canadian Journal of Research} \textbf{1940}, \emph{18},
  74--82\relax
\mciteBstWouldAddEndPuncttrue
\mciteSetBstMidEndSepPunct{\mcitedefaultmidpunct}
{\mcitedefaultendpunct}{\mcitedefaultseppunct}\relax
\EndOfBibitem
\bibitem[Phillips(1950)]{Phillips1950}
Phillips,~J.~G. On the identification of the 3670 \textup{\AA} band of the
  C$_2$ molecule. \emph{The Astrophysical Journal} \textbf{1950}, \emph{112},
  131\relax
\mciteBstWouldAddEndPuncttrue
\mciteSetBstMidEndSepPunct{\mcitedefaultmidpunct}
{\mcitedefaultendpunct}{\mcitedefaultseppunct}\relax
\EndOfBibitem
\bibitem[Messerle and Krauss(1967)Messerle, and Krauss]{Messerle1967}
Messerle,~G.; Krauss,~L. Ein neues $c'^1\pi_g-b^1\pi_u$-Bandensystem des
  C$_2$-Molek{\"u}ls. \emph{Z. Naturforsch.} \textbf{1967}, \emph{22 a},
  2015--2023\relax
\mciteBstWouldAddEndPuncttrue
\mciteSetBstMidEndSepPunct{\mcitedefaultmidpunct}
{\mcitedefaultendpunct}{\mcitedefaultseppunct}\relax
\EndOfBibitem
\bibitem[Anti{\'c}-Jovanovi{\'c} \latin{et~al.}(1985)Anti{\'c}-Jovanovi{\'c},
  Bojovi{\'c}, Pesi{\'c}, Vujisi{\'c}, Rakotoarijimy, and Weniger]{Antic1985}
Anti{\'c}-Jovanovi{\'c},~A.; Bojovi{\'c},~V.; Pesi{\'c},~D.; Vujisi{\'c},~B.;
  Rakotoarijimy,~D.; Weniger,~S. Study of isotopic $^{13}$C$_2$ bands of the
  Deslandres-d'Azambuja system. \emph{Journal of Molecular Spectroscopy}
  \textbf{1985}, \emph{110}, 86--92\relax
\mciteBstWouldAddEndPuncttrue
\mciteSetBstMidEndSepPunct{\mcitedefaultmidpunct}
{\mcitedefaultendpunct}{\mcitedefaultseppunct}\relax
\EndOfBibitem
\bibitem[Jiang \latin{et~al.}(2018)Jiang, Saladrigas, Erickson, Keenan, and
  Field]{jiang2018probing}
Jiang,~J.; Saladrigas,~C.~A.; Erickson,~T.~J.; Keenan,~C.~L.; Field,~R.~W.
  Probing the predissociated levels of the S$_1$ state of acetylene via H-atom
  fluorescence and photofragment fluorescence action spectroscopy. \emph{The
  Journal of Chemical Physics} \textbf{2018}, \emph{149}, 174309\relax
\mciteBstWouldAddEndPuncttrue
\mciteSetBstMidEndSepPunct{\mcitedefaultmidpunct}
{\mcitedefaultendpunct}{\mcitedefaultseppunct}\relax
\EndOfBibitem
\bibitem[Jiang \latin{et~al.}(2019)Jiang, Muthike, Erickson, and
  Field]{jiang2019one}
Jiang,~J.; Muthike,~A.~K.; Erickson,~T.~J.; Field,~R.~W. One-color (212--220
  nm) resonantly-enhanced (S$_1$--S$_0$) multi-photon dissociation of
  acetylene. \emph{Journal of Molecular Spectroscopy} \textbf{2019},
  \emph{361}, 24--33\relax
\mciteBstWouldAddEndPuncttrue
\mciteSetBstMidEndSepPunct{\mcitedefaultmidpunct}
{\mcitedefaultendpunct}{\mcitedefaultseppunct}\relax
\EndOfBibitem
\bibitem[Jiang \latin{et~al.}(2020)Jiang, Du, Li{\'e}vin, and
  Field]{jiang2020one}
Jiang,~J.; Du,~Z.; Li{\'e}vin,~J.; Field,~R.~W. One-colour ($\sim$220 nm)
  resonance-enhanced (S$_1$--S$_0$) multi-photon dissociation of acetylene:
  Probe of the C$_2$ $A^1\Pi_u-X^1\Sigma_g^+$ band by frequency-modulation
  spectroscopy. \emph{Molecular Physics} \textbf{2020}, \emph{118},
  e1724340\relax
\mciteBstWouldAddEndPuncttrue
\mciteSetBstMidEndSepPunct{\mcitedefaultmidpunct}
{\mcitedefaultendpunct}{\mcitedefaultseppunct}\relax
\EndOfBibitem
\bibitem[Ballik and Ramsay(1963)Ballik, and Ramsay]{Ballik1963}
Ballik,~E.; Ramsay,~D. An Extension of the Phillips system of C$_2$ and a
  survey of C$_2$ states. \emph{The Astrophysical Journal} \textbf{1963},
  \emph{137}, 84\relax
\mciteBstWouldAddEndPuncttrue
\mciteSetBstMidEndSepPunct{\mcitedefaultmidpunct}
{\mcitedefaultendpunct}{\mcitedefaultseppunct}\relax
\EndOfBibitem
\bibitem[Herzberg(1950)]{herzberg1950molecular}
Herzberg,~G. \emph{Molecular spectra and molecular structure. Vol. 1: Spectra
  of diatomic molecules}; New York: Van Nostrand Reinhold, 1950\relax
\mciteBstWouldAddEndPuncttrue
\mciteSetBstMidEndSepPunct{\mcitedefaultmidpunct}
{\mcitedefaultendpunct}{\mcitedefaultseppunct}\relax
\EndOfBibitem
\bibitem[Lefebvre-Brion and Field(2004)Lefebvre-Brion, and Field]{Bob2004}
Lefebvre-Brion,~H.; Field,~R.~W. \emph{The spectra and dynamics of diatomic
  molecules}; Elsevier Academic Press, 2004\relax
\mciteBstWouldAddEndPuncttrue
\mciteSetBstMidEndSepPunct{\mcitedefaultmidpunct}
{\mcitedefaultendpunct}{\mcitedefaultseppunct}\relax
\EndOfBibitem
\bibitem[McKemmish \latin{et~al.}(2020)McKemmish, Syme, Borsovszky, Yurchenko,
  Tennyson, Furtenbacher, and Cs{\'a}sz{\'a}r]{mckemmish2020update}
McKemmish,~L.~K.; Syme,~A.-M.; Borsovszky,~J.; Yurchenko,~S.~N.; Tennyson,~J.;
  Furtenbacher,~T.; Cs{\'a}sz{\'a}r,~A.~G. An update to the MARVEL data set and
  ExoMol line list for $^{12}$C$_2$. \emph{Monthly Notices of the Royal
  Astronomical Society} \textbf{2020}, \emph{497}, 1081--1097\relax
\mciteBstWouldAddEndPuncttrue
\mciteSetBstMidEndSepPunct{\mcitedefaultmidpunct}
{\mcitedefaultendpunct}{\mcitedefaultseppunct}\relax
\EndOfBibitem
\bibitem[Colbert and Miller(1992)Colbert, and Miller]{colbert1992novel}
Colbert,~D.~T.; Miller,~W.~H. A novel discrete variable representation for
  quantum mechanical reactive scattering via the S-matrix Kohn method.
  \emph{The Journal of Chemical Physics} \textbf{1992}, \emph{96},
  1982--1991\relax
\mciteBstWouldAddEndPuncttrue
\mciteSetBstMidEndSepPunct{\mcitedefaultmidpunct}
{\mcitedefaultendpunct}{\mcitedefaultseppunct}\relax
\EndOfBibitem
\bibitem[Chan and Head-Gordon(2002)Chan, and Head-Gordon]{chan2002highly}
Chan,~G. K.-L.; Head-Gordon,~M. Highly correlated calculations with a
  polynomial cost algorithm: A study of the density matrix renormalization
  group. \emph{The Journal of Chemical Physics} \textbf{2002}, \emph{116},
  4462--4476\relax
\mciteBstWouldAddEndPuncttrue
\mciteSetBstMidEndSepPunct{\mcitedefaultmidpunct}
{\mcitedefaultendpunct}{\mcitedefaultseppunct}\relax
\EndOfBibitem
\bibitem[Chan(2004)]{chan2004algorithm}
Chan,~G. K.-L. An algorithm for large scale density matrix renormalization
  group calculations. \emph{The Journal of Chemical Physics} \textbf{2004},
  \emph{120}, 3172--3178\relax
\mciteBstWouldAddEndPuncttrue
\mciteSetBstMidEndSepPunct{\mcitedefaultmidpunct}
{\mcitedefaultendpunct}{\mcitedefaultseppunct}\relax
\EndOfBibitem
\bibitem[Ghosh \latin{et~al.}(2008)Ghosh, Hachmann, Yanai, and
  Chan]{ghosh2008orbital}
Ghosh,~D.; Hachmann,~J.; Yanai,~T.; Chan,~G. K.-L. Orbital optimization in the
  density matrix renormalization group, with applications to polyenes and
  $\beta$-carotene. \emph{The Journal of Chemical Physics} \textbf{2008},
  \emph{128}, 144117\relax
\mciteBstWouldAddEndPuncttrue
\mciteSetBstMidEndSepPunct{\mcitedefaultmidpunct}
{\mcitedefaultendpunct}{\mcitedefaultseppunct}\relax
\EndOfBibitem
\bibitem[Sharma and Chan(2012)Sharma, and Chan]{sharma2012spin}
Sharma,~S.; Chan,~G. K.-L. Spin-adapted density matrix renormalization group
  algorithms for quantum chemistry. \emph{The Journal of Chemical Physics}
  \textbf{2012}, \emph{136}, 124121\relax
\mciteBstWouldAddEndPuncttrue
\mciteSetBstMidEndSepPunct{\mcitedefaultmidpunct}
{\mcitedefaultendpunct}{\mcitedefaultseppunct}\relax
\EndOfBibitem
\bibitem[Olivares-Amaya \latin{et~al.}(2015)Olivares-Amaya, Hu, Nakatani,
  Sharma, Yang, and Chan]{olivares2015ab}
Olivares-Amaya,~R.; Hu,~W.; Nakatani,~N.; Sharma,~S.; Yang,~J.; Chan,~G. K.-L.
  The ab-initio density matrix renormalization group in practice. \emph{The
  Journal of Chemical Physics} \textbf{2015}, \emph{142}, 034102\relax
\mciteBstWouldAddEndPuncttrue
\mciteSetBstMidEndSepPunct{\mcitedefaultmidpunct}
{\mcitedefaultendpunct}{\mcitedefaultseppunct}\relax
\EndOfBibitem
\bibitem[Sun \latin{et~al.}(2018)Sun, Berkelbach, Blunt, Booth, Guo, Li, Liu,
  McClain, Sayfutyarova, Sharma, \latin{et~al.} others]{sun2018pyscf}
Sun,~Q.; Berkelbach,~T.~C.; Blunt,~N.~S.; Booth,~G.~H.; Guo,~S.; Li,~Z.;
  Liu,~J.; McClain,~J.~D.; Sayfutyarova,~E.~R.; Sharma,~S., \latin{et~al.}
  PySCF: the Python-based simulations of chemistry framework. \emph{Wiley
  Interdisciplinary Reviews: Computational Molecular Science} \textbf{2018},
  \emph{8}, e1340\relax
\mciteBstWouldAddEndPuncttrue
\mciteSetBstMidEndSepPunct{\mcitedefaultmidpunct}
{\mcitedefaultendpunct}{\mcitedefaultseppunct}\relax
\EndOfBibitem
\bibitem[Schmidt(2021)]{schmidt2021spectroscopy}
Schmidt,~T.~W. The Spectroscopy of C$_2$: A Cosmic Beacon. \emph{Accounts of
  Chemical Research} \textbf{2021}, \emph{54}, 481--489\relax
\mciteBstWouldAddEndPuncttrue
\mciteSetBstMidEndSepPunct{\mcitedefaultmidpunct}
{\mcitedefaultendpunct}{\mcitedefaultseppunct}\relax
\EndOfBibitem
\bibitem[Werner \latin{et~al.}(2020)Werner, Knowles, Manby, Black, Doll,
  He{\ss}elmann, Kats, K{\"o}hn, Korona, Kreplin, \latin{et~al.}
  others]{werner2020molpro}
Werner,~H.-J.; Knowles,~P.~J.; Manby,~F.~R.; Black,~J.~A.; Doll,~K.;
  He{\ss}elmann,~A.; Kats,~D.; K{\"o}hn,~A.; Korona,~T.; Kreplin,~D.~A.,
  \latin{et~al.}  The Molpro quantum chemistry package. \emph{The Journal of
  Chemical Physics} \textbf{2020}, \emph{152}, 144107\relax
\mciteBstWouldAddEndPuncttrue
\mciteSetBstMidEndSepPunct{\mcitedefaultmidpunct}
{\mcitedefaultendpunct}{\mcitedefaultseppunct}\relax
\EndOfBibitem
\bibitem[Krechkivska \latin{et~al.}(2017)Krechkivska, Welsh, Bacskay, Nauta,
  Kable, and Schmidt]{krechkivska2017first}
Krechkivska,~O.; Welsh,~B.; Bacskay,~G.; Nauta,~K.; Kable,~S.; Schmidt,~T.
  First observation of the $3^3\Pi_g$ state of C$_2$: Born-Oppenheimer
  breakdown. \emph{The Journal of Chemical Physics} \textbf{2017}, \emph{146},
  134306\relax
\mciteBstWouldAddEndPuncttrue
\mciteSetBstMidEndSepPunct{\mcitedefaultmidpunct}
{\mcitedefaultendpunct}{\mcitedefaultseppunct}\relax
\EndOfBibitem
\bibitem[Neese \latin{et~al.}(2020)Neese, Wennmohs, Becker, and
  Riplinger]{neese2020orca}
Neese,~F.; Wennmohs,~F.; Becker,~U.; Riplinger,~C. The ORCA quantum chemistry
  program package. \emph{The Journal of Chemical Physics} \textbf{2020},
  \emph{152}, 224108\relax
\mciteBstWouldAddEndPuncttrue
\mciteSetBstMidEndSepPunct{\mcitedefaultmidpunct}
{\mcitedefaultendpunct}{\mcitedefaultseppunct}\relax
\EndOfBibitem
\bibitem[Phillips(1949)]{phillips1949fox}
Phillips,~J.~G. The Fox-Herzberg system of the C$_2$ molecule. \emph{The
  Astrophysical Journal} \textbf{1949}, \emph{110}, 73\relax
\mciteBstWouldAddEndPuncttrue
\mciteSetBstMidEndSepPunct{\mcitedefaultmidpunct}
{\mcitedefaultendpunct}{\mcitedefaultseppunct}\relax
\EndOfBibitem
\bibitem[Jia \latin{et~al.}(2012)Jia, Diao, Liu, Wang, Liu, and
  Zhang]{jia2012equivalence}
Jia,~C.-S.; Diao,~Y.-F.; Liu,~X.-J.; Wang,~P.-Q.; Liu,~J.-Y.; Zhang,~G.-D.
  Equivalence of the Wei potential model and Tietz potential model for diatomic
  molecules. \emph{The Journal of Chemical Physics} \textbf{2012}, \emph{137},
  014101\relax
\mciteBstWouldAddEndPuncttrue
\mciteSetBstMidEndSepPunct{\mcitedefaultmidpunct}
{\mcitedefaultendpunct}{\mcitedefaultseppunct}\relax
\EndOfBibitem
\bibitem[Hua(1990)]{hua1990four}
Hua,~W. Four-parameter exactly solvable potential for diatomic molecules.
  \emph{Physical Review A} \textbf{1990}, \emph{42}, 2524\relax
\mciteBstWouldAddEndPuncttrue
\mciteSetBstMidEndSepPunct{\mcitedefaultmidpunct}
{\mcitedefaultendpunct}{\mcitedefaultseppunct}\relax
\EndOfBibitem
\bibitem[Tietz(1963)]{tietz1963potential}
Tietz,~T. Potential-energy function for diatomic molecules. \emph{The Journal
  of Chemical Physics} \textbf{1963}, \emph{38}, 3036--3037\relax
\mciteBstWouldAddEndPuncttrue
\mciteSetBstMidEndSepPunct{\mcitedefaultmidpunct}
{\mcitedefaultendpunct}{\mcitedefaultseppunct}\relax
\EndOfBibitem
\bibitem[Borsovszky \latin{et~al.}(2021)Borsovszky, Nauta, Jiang, Hansen,
  McKemmish, Field, Stanton, Kable, and Schmidt]{Borsovszky2021}
Borsovszky,~J.; Nauta,~K.; Jiang,~J.; Hansen,~C.~S.; McKemmish,~L.~K.;
  Field,~R.~W.; Stanton,~J.~F.; Kable,~S.~H.; Schmidt,~T.~W. Photodissociation
  of dicarbon: How nature breaks an unusual multiple bond. \emph{Proceedings of
  the National Academy of Sciences of the United States of America}
  \textbf{2021}, \emph{118}, e2113315118\relax
\mciteBstWouldAddEndPuncttrue
\mciteSetBstMidEndSepPunct{\mcitedefaultmidpunct}
{\mcitedefaultendpunct}{\mcitedefaultseppunct}\relax
\EndOfBibitem
\bibitem[Chen \latin{et~al.}(2015)Chen, Kawaguchi, Bernath, and
  Tang]{chen2015simultaneous}
Chen,~W.; Kawaguchi,~K.; Bernath,~P.~F.; Tang,~J. Simultaneous analysis of the
  Ballik-Ramsay and Phillips systems of C$_2$ and observation of forbidden
  transitions between singlet and triplet states. \emph{The Journal of Chemical
  Physics} \textbf{2015}, \emph{142}, 064317\relax
\mciteBstWouldAddEndPuncttrue
\mciteSetBstMidEndSepPunct{\mcitedefaultmidpunct}
{\mcitedefaultendpunct}{\mcitedefaultseppunct}\relax
\EndOfBibitem
\bibitem[Huber and Herzberg(1979)Huber, and Herzberg]{Huber1979}
Huber,~K.-P.; Herzberg,~G. \emph{Molecular spectra and molecular structure: IV.
  Constants of diatomic molecules}; Van Nostrand Reinhold Company, 1979\relax
\mciteBstWouldAddEndPuncttrue
\mciteSetBstMidEndSepPunct{\mcitedefaultmidpunct}
{\mcitedefaultendpunct}{\mcitedefaultseppunct}\relax
\EndOfBibitem
\bibitem[Furtenbacher \latin{et~al.}(2016)Furtenbacher, Szab{\'o},
  Cs{\'a}sz{\'a}r, Bernath, Yurchenko, and
  Tennyson]{furtenbacher2016experimental}
Furtenbacher,~T.; Szab{\'o},~I.; Cs{\'a}sz{\'a}r,~A.~G.; Bernath,~P.~F.;
  Yurchenko,~S.~N.; Tennyson,~J. Experimental energy levels and partition
  function of the $^{12}$C$_2$ molecule. \emph{The Astrophysical Journal
  Supplement Series} \textbf{2016}, \emph{224}, 44\relax
\mciteBstWouldAddEndPuncttrue
\mciteSetBstMidEndSepPunct{\mcitedefaultmidpunct}
{\mcitedefaultendpunct}{\mcitedefaultseppunct}\relax
\EndOfBibitem
\bibitem[Lewis \latin{et~al.}(2005)Lewis, Gibson, Zhang, Lefebvre-Brion, and
  Robbe]{lewis2005predissociation}
Lewis,~B.; Gibson,~S.; Zhang,~W.; Lefebvre-Brion,~H.; Robbe,~J.-M.
  Predissociation mechanism for the lowest $^1\Pi_u$ states of N$_2$. \emph{The
  Journal of Chemical Physics} \textbf{2005}, \emph{122}, 144302\relax
\mciteBstWouldAddEndPuncttrue
\mciteSetBstMidEndSepPunct{\mcitedefaultmidpunct}
{\mcitedefaultendpunct}{\mcitedefaultseppunct}\relax
\EndOfBibitem
\bibitem[Lewis \latin{et~al.}(2008)Lewis, Heays, Gibson, Lefebvre-Brion, and
  Lefebvre]{lewis2008coupled}
Lewis,~B.; Heays,~A.; Gibson,~S.; Lefebvre-Brion,~H.; Lefebvre,~R. A
  coupled-channel model of the $^3\Pi_u$ states of N$_2$: Structure and
  interactions of the $3s\sigma_gF_3$ $^3\Pi_u$ and $3p\pi_uG_3$ $^3\Pi_u$
  Rydberg states. \emph{The Journal of Chemical Physics} \textbf{2008},
  \emph{129}, 164306\relax
\mciteBstWouldAddEndPuncttrue
\mciteSetBstMidEndSepPunct{\mcitedefaultmidpunct}
{\mcitedefaultendpunct}{\mcitedefaultseppunct}\relax
\EndOfBibitem
\bibitem[Krechkivska \latin{et~al.}(2015)Krechkivska, Bacskay, Troy, Nauta,
  Kreuscher, Kable, and Schmidt]{krechkivska2015resonance}
Krechkivska,~O.; Bacskay,~G.~B.; Troy,~T.~P.; Nauta,~K.; Kreuscher,~T.~D.;
  Kable,~S.~H.; Schmidt,~T.~W. Resonance-enhanced 2-photon ionization scheme
  for C$_2$ through a newly identified band system: $4^3\Pi_g-a^3\Pi_u$.
  \emph{The Journal of Physical Chemistry A} \textbf{2015}, \emph{119},
  12102--12108\relax
\mciteBstWouldAddEndPuncttrue
\mciteSetBstMidEndSepPunct{\mcitedefaultmidpunct}
{\mcitedefaultendpunct}{\mcitedefaultseppunct}\relax
\EndOfBibitem
\bibitem[Blunt \latin{et~al.}(1995)Blunt, Lin, Sorkhabi, and
  Jackson]{blunt1995revised}
Blunt,~V.; Lin,~H.; Sorkhabi,~O.; Jackson,~W. Revised molecular constants for
  the $D^1\Sigma_u^+$ state of C$_2$. \emph{Journal of Molecular Spectroscopy}
  \textbf{1995}, \emph{174}, 274--276\relax
\mciteBstWouldAddEndPuncttrue
\mciteSetBstMidEndSepPunct{\mcitedefaultmidpunct}
{\mcitedefaultendpunct}{\mcitedefaultseppunct}\relax
\EndOfBibitem
\bibitem[Krechkivska \latin{et~al.}(2018)Krechkivska, Welsh, Fr{\'e}reux,
  Nauta, Kable, and Schmidt]{krechkivska2018higher}
Krechkivska,~O.; Welsh,~B.; Fr{\'e}reux,~J.; Nauta,~K.; Kable,~S.; Schmidt,~T.
  Higher vibrational levels of the $D^1\Sigma_u^+$ state of dicarbon: New
  Mulliken bands. \emph{Journal of Molecular Spectroscopy} \textbf{2018},
  \emph{344}, 1--5\relax
\mciteBstWouldAddEndPuncttrue
\mciteSetBstMidEndSepPunct{\mcitedefaultmidpunct}
{\mcitedefaultendpunct}{\mcitedefaultseppunct}\relax
\EndOfBibitem
\bibitem[Joester \latin{et~al.}(2007)Joester, Nakajima, Reilly, Kokkin, Nauta,
  Kable, and Schmidt]{joester2007d}
Joester,~J.~A.; Nakajima,~M.; Reilly,~N.~J.; Kokkin,~D.~L.; Nauta,~K.;
  Kable,~S.~H.; Schmidt,~T.~W. The $d^3\Pi_g-c^3\Sigma_u^+$ band system of
  C$_2$. \emph{The Journal of Chemical Physics} \textbf{2007}, \emph{127},
  214303--214303\relax
\mciteBstWouldAddEndPuncttrue
\mciteSetBstMidEndSepPunct{\mcitedefaultmidpunct}
{\mcitedefaultendpunct}{\mcitedefaultseppunct}\relax
\EndOfBibitem
\bibitem[Chan \latin{et~al.}(2013)Chan, Yeung, Wang, and Cheung]{chan2013laser}
Chan,~M.-C.; Yeung,~S.-H.; Wang,~N.; Cheung,~A.-C. Laser absorption
  spectroscopy of the $d^3\Pi_g-c^3\Sigma_u^+$ transition of C$_2$. \emph{The
  Journal of Physical Chemistry A} \textbf{2013}, \emph{117}, 9578--9583\relax
\mciteBstWouldAddEndPuncttrue
\mciteSetBstMidEndSepPunct{\mcitedefaultmidpunct}
{\mcitedefaultendpunct}{\mcitedefaultseppunct}\relax
\EndOfBibitem
\bibitem[Nakajima and Endo(2013)Nakajima, and Endo]{nakajima2013spectroscopic}
Nakajima,~M.; Endo,~Y. Spectroscopic observation of higher vibrational levels
  of C$_2$ through visible band systems. \emph{The Journal of Chemical Physics}
  \textbf{2013}, \emph{139}, 244310\relax
\mciteBstWouldAddEndPuncttrue
\mciteSetBstMidEndSepPunct{\mcitedefaultmidpunct}
{\mcitedefaultendpunct}{\mcitedefaultseppunct}\relax
\EndOfBibitem
\bibitem[Nakajima and Endo(2014)Nakajima, and Endo]{nakajima2014excitation}
Nakajima,~M.; Endo,~Y. Excitation spectra of the $d^3\Pi_g-c^3\Sigma_u^+$ band
  system of C$_2$. \emph{Journal of Molecular Spectroscopy} \textbf{2014},
  \emph{302}, 9--16\relax
\mciteBstWouldAddEndPuncttrue
\mciteSetBstMidEndSepPunct{\mcitedefaultmidpunct}
{\mcitedefaultendpunct}{\mcitedefaultseppunct}\relax
\EndOfBibitem
\bibitem[Shaik \latin{et~al.}(2016)Shaik, Danovich, Braida, and
  Hiberty]{shaik2016quadruple}
Shaik,~S.; Danovich,~D.; Braida,~B.; Hiberty,~P.~C. The quadruple bonding in
  C$_2$ reproduces the properties of the molecule. \emph{Chemistry-A European
  Journal} \textbf{2016}, \emph{22}, 4116--4128\relax
\mciteBstWouldAddEndPuncttrue
\mciteSetBstMidEndSepPunct{\mcitedefaultmidpunct}
{\mcitedefaultendpunct}{\mcitedefaultseppunct}\relax
\EndOfBibitem
\bibitem[Su \latin{et~al.}(2011)Su, Wu, Gu, Wu, Shaik, and
  Hiberty]{su2011bonding}
Su,~P.; Wu,~J.; Gu,~J.; Wu,~W.; Shaik,~S.; Hiberty,~P.~C. Bonding conundrums in
  the C$_2$ molecule: A valence bond study. \emph{Journal of Chemical Theory
  and Computation} \textbf{2011}, \emph{7}, 121--130\relax
\mciteBstWouldAddEndPuncttrue
\mciteSetBstMidEndSepPunct{\mcitedefaultmidpunct}
{\mcitedefaultendpunct}{\mcitedefaultseppunct}\relax
\EndOfBibitem
\bibitem[Abrams and Sherrill(2004)Abrams, and Sherrill]{abrams2004full}
Abrams,~M.~L.; Sherrill,~C.~D. Full configuration interaction potential energy
  curves for the $X^1\Sigma_g^+$, $B^1\Delta_g$, and $B'^1\Sigma_g^+$ states of
  C$_2$: A challenge for approximate methods. \emph{The Journal of Chemical
  Physics} \textbf{2004}, \emph{121}, 9211--9219\relax
\mciteBstWouldAddEndPuncttrue
\mciteSetBstMidEndSepPunct{\mcitedefaultmidpunct}
{\mcitedefaultendpunct}{\mcitedefaultseppunct}\relax
\EndOfBibitem
\bibitem[Varandas(2008)]{varandas2008extrapolation}
Varandas,~A. Extrapolation to the complete-basis-set limit and the implications
  of avoided crossings: The $X^1\Sigma_g^+$, $B^1\Delta_g$, and $B'\Sigma_g^+$
  states of C$_2$. \emph{The Journal of Chemical Physics} \textbf{2008},
  \emph{129}, 234103\relax
\mciteBstWouldAddEndPuncttrue
\mciteSetBstMidEndSepPunct{\mcitedefaultmidpunct}
{\mcitedefaultendpunct}{\mcitedefaultseppunct}\relax
\EndOfBibitem
\bibitem[Douay \latin{et~al.}(1988)Douay, Nietmann, and
  Bernath]{douay1988discovery}
Douay,~M.; Nietmann,~R.; Bernath,~P. The discovery of two new infrared
  electronic transitions of C$_2$: $B^1\Delta_g-A^1\Pi_u$ and
  $B'^1\Sigma_g^+-A^1\Pi_u$. \emph{Journal of Molecular Spectroscopy}
  \textbf{1988}, \emph{131}, 261--271\relax
\mciteBstWouldAddEndPuncttrue
\mciteSetBstMidEndSepPunct{\mcitedefaultmidpunct}
{\mcitedefaultendpunct}{\mcitedefaultseppunct}\relax
\EndOfBibitem
\bibitem[Zewail(2000)]{zewail2000femtochemistry}
Zewail,~A.~H. Femtochemistry: Atomic-scale dynamics of the chemical bond using
  ultrafast lasers (Nobel Lecture). \emph{Angewandte Chemie International
  Edition} \textbf{2000}, \emph{39}, 2586--2631\relax
\mciteBstWouldAddEndPuncttrue
\mciteSetBstMidEndSepPunct{\mcitedefaultmidpunct}
{\mcitedefaultendpunct}{\mcitedefaultseppunct}\relax
\EndOfBibitem
\bibitem[Welsh \latin{et~al.}(2017)Welsh, Krechkivska, Nauta, Bacskay, Kable,
  and Schmidt]{welsh2017pi}
Welsh,~B.; Krechkivska,~O.; Nauta,~K.; Bacskay,~G.; Kable,~S.; Schmidt,~T. The
  $e^3\Pi_g$ state of C$_2$: A pathway to dissociation. \emph{The Journal of
  Chemical Physics} \textbf{2017}, \emph{147}, 024305\relax
\mciteBstWouldAddEndPuncttrue
\mciteSetBstMidEndSepPunct{\mcitedefaultmidpunct}
{\mcitedefaultendpunct}{\mcitedefaultseppunct}\relax
\EndOfBibitem
\bibitem[Gelbart and Freed(1973)Gelbart, and Freed]{gelbart1973intramolecular}
Gelbart,~W.~M.; Freed,~K.~F. Intramolecular perturbations and the quenching of
  luminescence in small molecules. \emph{Chemical Physics Letters}
  \textbf{1973}, \emph{18}, 470--475\relax
\mciteBstWouldAddEndPuncttrue
\mciteSetBstMidEndSepPunct{\mcitedefaultmidpunct}
{\mcitedefaultendpunct}{\mcitedefaultseppunct}\relax
\EndOfBibitem
\end{mcitethebibliography}


\providecommand{\latin}[1]{#1}
\makeatletter
\providecommand{\doi}
  {\begingroup\let\do\@makeother\dospecials
  \catcode`\{=1 \catcode`\}=2 \doi@aux}
\providecommand{\doi@aux}[1]{\endgroup\texttt{#1}}
\makeatother
\providecommand*\mcitethebibliography{\thebibliography}
\csname @ifundefined\endcsname{endmcitethebibliography}
  {\let\endmcitethebibliography\endthebibliography}{}
\begin{mcitethebibliography}{6}
\providecommand*\natexlab[1]{#1}
\providecommand*\mciteSetBstSublistMode[1]{}
\providecommand*\mciteSetBstMaxWidthForm[2]{}
\providecommand*\mciteBstWouldAddEndPuncttrue
  {\def\EndOfBibitem{\unskip.}}
\providecommand*\mciteBstWouldAddEndPunctfalse
  {\let\EndOfBibitem\relax}
\providecommand*\mciteSetBstMidEndSepPunct[3]{}
\providecommand*\mciteSetBstSublistLabelBeginEnd[3]{}
\providecommand*\EndOfBibitem{}
\mciteSetBstSublistMode{f}
\mciteSetBstMaxWidthForm{subitem}{(\alph{mcitesubitemcount})}
\mciteSetBstSublistLabelBeginEnd
  {\mcitemaxwidthsubitemform\space}
  {\relax}
  {\relax}

\bibitem[Messerle and Krauss(1967)Messerle, and Krauss]{Messerle1967}
Messerle,~G.; Krauss,~L. Ein neues $c'^1\pi_g-b^1\pi_u$-Bandensystem des
  C$_2$-Molek{\"u}ls. \emph{Z. Naturforsch.} \textbf{1967}, \emph{22 a},
  2015--2023\relax
\mciteBstWouldAddEndPuncttrue
\mciteSetBstMidEndSepPunct{\mcitedefaultmidpunct}
{\mcitedefaultendpunct}{\mcitedefaultseppunct}\relax
\EndOfBibitem
\bibitem[Messerle(1967)]{messerlethesis}
Messerle,~G. Eine Untersuchung der $\Pi_g$-Terme zur spektroskopischen
  Bestimmung der Dissoziationsenergie des \textup{C}$_2$-Moleküls. Ph.D.\
  thesis, Technische Hochschule München, 1967\relax
\mciteBstWouldAddEndPuncttrue
\mciteSetBstMidEndSepPunct{\mcitedefaultmidpunct}
{\mcitedefaultendpunct}{\mcitedefaultseppunct}\relax
\EndOfBibitem
\bibitem[Dieke and Lochte-Holtgreven(1930)Dieke, and
  Lochte-Holtgreven]{Dieke1930}
Dieke,~G.; Lochte-Holtgreven,~W. {\"U}ber einige Banden des
  Kohlenstoffimolek{\"u}ls. \emph{Zeitschrift f{\"u}r Physik} \textbf{1930},
  \emph{62}, 767--794\relax
\mciteBstWouldAddEndPuncttrue
\mciteSetBstMidEndSepPunct{\mcitedefaultmidpunct}
{\mcitedefaultendpunct}{\mcitedefaultseppunct}\relax
\EndOfBibitem
\bibitem[Herzberg and Sutton(1940)Herzberg, and Sutton]{Herzberg1940}
Herzberg,~G.; Sutton,~R. Tail bands of the Deslandres-d'Azambuja system of the
  C$_2$ molecule. \emph{Canadian Journal of Research} \textbf{1940}, \emph{18},
  74--82\relax
\mciteBstWouldAddEndPuncttrue
\mciteSetBstMidEndSepPunct{\mcitedefaultmidpunct}
{\mcitedefaultendpunct}{\mcitedefaultseppunct}\relax
\EndOfBibitem
\bibitem[Phillips(1950)]{Phillips1950}
Phillips,~J.~G. On the identification of the 3670 \textup{\AA} band of the
  C$_2$ molecule. \emph{The Astrophysical Journal} \textbf{1950}, \emph{112},
  131\relax
\mciteBstWouldAddEndPuncttrue
\mciteSetBstMidEndSepPunct{\mcitedefaultmidpunct}
{\mcitedefaultendpunct}{\mcitedefaultseppunct}\relax
\EndOfBibitem
\end{mcitethebibliography}

\end{document}




\newpage
\section{Non-existence of the Messerle and Krauss $C'^1\Pi_g$ state}
\label{sec:MK}

The assignment of the $C'^1\Pi_g$ states by Messerle and Krauss~\cite{Messerle1967} is based on the observed small anomalies in the $C$-state rotational combination differences, $\Delta_2 F(J)=F(J+1)-F(J-1)$. For a non-rigid rotor described by Eq.~6, 
\begin{equation} \label{d2F}
\Delta_2 F(J)/(J+1/2)=(4B_v+2D_v)-8D_v(J+1/2)^2.
\end{equation}
By plotting $\Delta_2 F(J)/(J+1/2)$ $vs.$ $(J+1/2)^2$ (referred to as ``the $\Delta_2 F(J)$ plot''), one obtains a straight line with a negative slope of -$8D_v$, and a vertical-intercept at $4B_v+2D_v$. Higher-order centrifugal distortion corrections to Eq.~6 (e.g. $H_v[J(J+1)]^3$) are not expected to significantly alter the $\Delta_2F(J)$ plot, given that these higher-order terms are expected to be several orders of magnitude smaller than $D_v[J(J+1)-1]^2$. Significant deviations from a straight line in the $\Delta_2 F(J)$ plot are therefore indications of perturbations. 


\begin{figure}[t]
\center
\includegraphics[width=4.75 in]{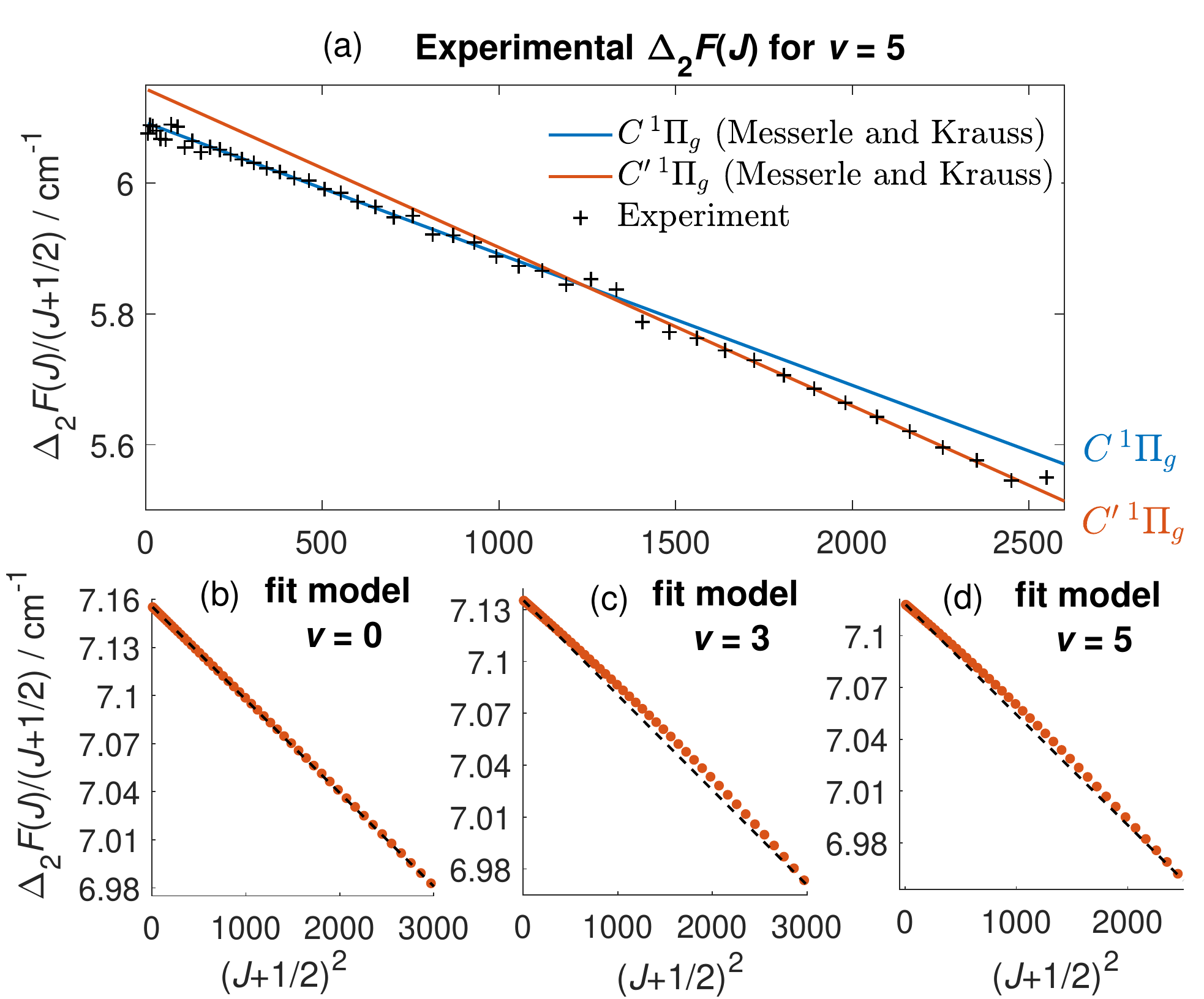}
\caption{The $\Delta_2 F(J)$ plots. (a) The experimental plot for the $v=5$ level of the $C^1\Pi_g$ state, based on the line-list for the $C^1\Pi_g (v=5)-A^1\Pi_u (v=4)$ band transitions in Ref.~\citenum{messerlethesis}. (b)-(d) The calculated plots (orange dots) for the $C$-state $v=0$, 3, and 5 levels from the B-R two-state model. The first and last data points in panels (b)-(d) are connected by a straight dashed line to assist visualization of the curvatures in $\Delta_2 F(J)$.} 
\label{fig:MK_curv}
\end{figure}

According to Messerle and Krauss~\cite{Messerle1967}, perturbation signatures are observed in the $\Delta_2 F(J)$ plots for the $C$-state $v=3-8$ levels, but not in similar plots for $v=0-2$. The $\Delta_2 F(J)$ plot for the $v=5$ level is shown in Fig.~\ref{fig:MK_curv}a. As is shown by the two color-coded straight lines, the slope of the experimental $\Delta_2 F(J)$ plot is $J$-dependent. The $\Delta_2 F(J)$ plot at high-$J$ appears to have a slightly steeper negative slope than at low-$J$. The two color-coded lines in Fig.~\ref{fig:MK_curv}a intersect at $J\sim36$. Messerle and Krauss assigned the low-$J$ levels in the $\Delta_2 F(J)$ plot (i.e. $J\lesssim36$) to the $C^1\Pi_g$ $v=5$ level, and the higher $J$ levels to $v=2$ of their proposed $C'^1\Pi_g$ perturber state. Using a similar assignment scheme, Messerle and Krauss concluded that the $C$-state $v=n$ level (with $n\geq3$) is locally perturbed by the $C'$-state $v=n-3$ level (see Fig.~3b of the main text).  Given the experimental line position uncertainties of 0.2 cm$^{-1}$, these small anomalies in the $\Delta_2 F(J)$ plots only become noticeable after the inclusion of high-$J$ term values ($J\sim50$). These high-$J$ levels were observed only in the work of Messerle and Krauss~\cite{Messerle1967,messerlethesis}, which included only an experimental line-list for the $C^1\Pi_g (v=5)-A^1\Pi_u (v=4)$ band transitions~\cite{messerlethesis}. 

We disagree with this assignment scheme of Messerle and Krauss, which leads to unphysical consequences. For example, based on their assignments for levels relevant to the $\Delta_2 F(J)$ plot in Fig.~\ref{fig:MK_curv}a, at low $J$ (i.e. prior to the assumed $C/C'$ level-crossing), the $C$ state is the only spectroscopically bright state, since the low-$J$ experimental observations fall exclusively on the blue line, which represents the zeroth-order $C$-state values. The $C'-A$ oscillator strength appears to be negligible at low $J$ values. However, the brightness of the $C-A$ transition appears to abruptly disappear at $J\sim37$. At higher $J$, the $C'$ state becomes the only bright state. There is no physical mechanism that could cause this sudden switch of the spectroscopic transition brightness from $C-A$ to $C'-A$. 

The perturbation analysis of Messerle and Krauss was based on incorrect interpretations of the observed $\Delta_2 F(J)$ plots. Based on our two-state model, both the low- and high-$J$ levels in a given $\Delta_2 F(J)$ plot are firmly assigned to one specific vibrational level of the $C^1\Pi_g$ state. As is evident in Figs.~\ref{fig:MK_curv}b-\ref{fig:MK_curv}d, the observed patterns in the $\Delta_2 F(J)$ plots reported by Messerle and Krauss are reproduced by our fit model. For example, the calculated $\Delta_2 F(J)$ plot for the $v=0$ level (Fig.~\ref{fig:MK_curv}b) is close to a straight line, consistent with experimental observations. In higher-energy vibrational levels, for which apparent changes of slopes were observed in the experimental $\Delta_2 F(J)$ plots, curvatures in the calculated $\Delta_2 F(J)$ plots are also obvious (e.g. the $v=3$ and 5 levels in Figs.~\ref{fig:MK_curv}c and \ref{fig:MK_curv}d). An arbitrary separation of the $\Delta_2 F(J)$ plot into two distinct sections leads to the incorrect conclusion from Messerle and Krauss about the existence of the local $C'^1\Pi_g$ perturber for each of the $C$-state vibrational levels with $v\ge3$. %

We conclude that the assignment of this local $C'$-state perturber by Messerle and Krauss is based on unphysical conclusions. Foremost among these is an abrupt transfer of the spectroscopic transition brightness from the $C$ state to its assumed $C'$-state perturber that occurs in a narrow $J$-window, as well as a physically impossible avoided-crossing pattern between the $C$- and $C'$-state PECs (see Fig. 3b and the discussion in the main text). The so-called
Messerle-Krauss band system ($C'^1\Pi_g-A^1\Pi_u$) has been incorrectly assigned. The Messerle and Krauss $C'^1\Pi_g$ state does not exist.

\section{Valence-hole two-state fit model of the C$_2$ $C^1\Pi_g$ state}
\label{sec:numericals_vh}

\begin{table} 
\caption{Molecular parameters derived for the valence-hole two-state fit model for the C$_2$ $C^1\Pi_g$ state. Numbers in parentheses are $1\sigma$ uncertainties of the last digits. The asterisks indicate parameters with fixed values. The listed $T_e$ values are the energies at the minima of the two diabatic potentials relative to the minimum of the C$_2$ $X$-state potential. The $T_e$ values relative to the energy of the 2s$^2$2p$^2$($^3P$)+2s$^2$2p$^2$($^3P$) limit (51315 cm$^{-1}$) are given in the last column, $T_e$\,(Diss). The 2s$^2$2p$^2$ electron configuration label is omitted for the $^3P$ term in the next to last column that specifies the dissociation limits for the two diabats.}
\begin{center}
\begin{tabular}{@{\hspace{8pt}} c @{\hspace{8pt}} | @{\hspace{6pt}} c @{\hspace{6pt}}  @{\hspace{6pt}} c   @{\hspace{6pt}}c  @{\hspace{6pt}} c  @{\hspace{6pt}}  @{\hspace{6pt}} c   @{\hspace{6pt}}c   @{\hspace{6pt}} c @{\hspace{6pt}} @{\hspace{6pt}} c   @{\hspace{6pt}}c  | @{\hspace{6pt}} c @{\hspace{6pt}} | @{\hspace{6pt}} c @{\hspace{6pt}}}

\hline
diabats & $T_e$\,/\,cm$^{-1}$ & $\beta$\,/\,$\text{\AA}^{-1}$ & & $R_e\,/\,\text{\AA}$& $h$* & Diss. Limit& $T_e$\,(Diss)\,/\,cm$^{-1}$\\
\hline
d$_1$ & 35651.9(198) & 2.1604(28) & &	1.2461(4) & 0.73 & ($^3P$)+2s2p$^3$($^3D$)&-15663.1(198)\\
d$_2$& 47352.4(1747) & 4.3895(398)& & 1.6519(60)& 225 & ($^3P$)+($^3P$) & -3962.6(1747)\\
\hline
 & $H_e$\,/\,cm$^{-1}$ & && &&\\
\hline
d$_1$-d$_2$ & 7256.1(2258) & && &&\\

\end{tabular}
\end{center}
\label{tab:valence_hole_two_state}
\end{table}

The derived fit parameters for the valence-hole two-state model are listed in Table~\ref{tab:valence_hole_two_state}. The average fit residual for this fit model is 0.4 cm$^{-1}$, which is significantly smaller than that of the B-R model (2 cm$^{-1}$, see the main text). Unlike the B-R model, the four-parameter potential function of Eq.~7 is used for both the d$_1$ and d$_2$ diabats. Other aspects of the valence-hole two-state fit model (e.g. use of an $R$-independent electronic matrix element, $H_{12}^{el}$ ) are identical to those of the B-R fit model described in the Methods Section of the main text. 

\begin{figure}[t]
\center
\includegraphics[width=4.8 in]{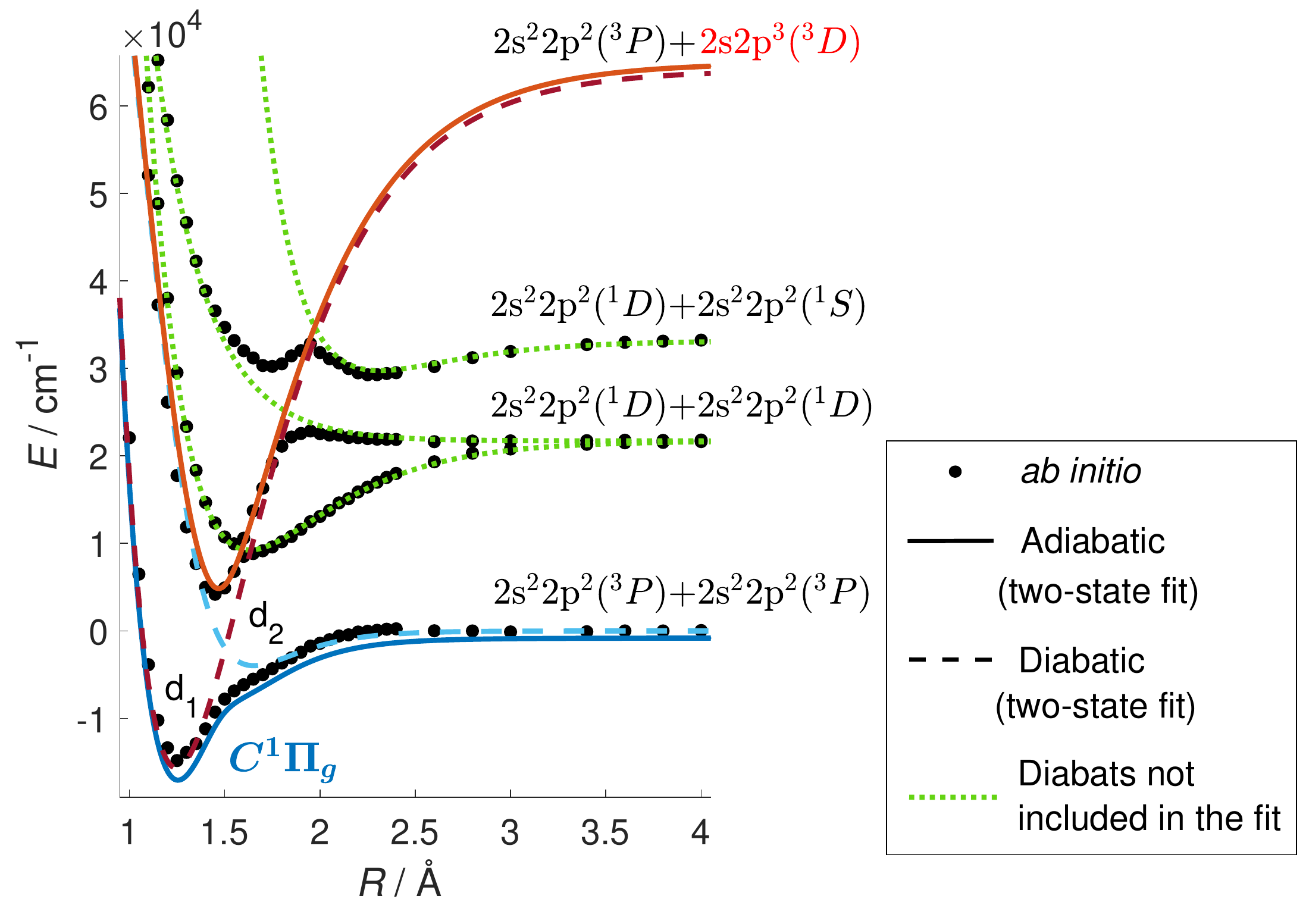}
\caption{The $^1\Pi_g$ PECs from the valence-hole two-state (d$_1$ and d$_2$) fit model and the $ab$ $initio$ calculation, obtained with the DMRG method. The green dotted lines indicate higher-energy diabats that are not included in the effective two-state fit.}
\label{fig:vh_two_state_fit}
\end{figure}

The PECs from our valence-hole fit model are shown in Fig.~\ref{fig:vh_two_state_fit}, along with the $ab$ $initio$ results from the DMRG calculation. The B-R model (see Fig.~3 of the main text) predicts that the minimum of the second adiabatic $^1\Pi_g$ potential lies slightly below the energy of the 2s$^2$2p$^2$($^3P$)+2s$^2$2p$^2$($^3P$) threshold. However, this minimum energy is predicted to be $\sim$5000 cm$^{-1}$ above that threshold by the DMRG calculation. As is evident from Fig.~\ref{fig:vh_two_state_fit}, the result from the valence-hole two-state fit model is in excellent agreement with the $ab$ $initio$ calculation regarding the equilibrium energy of this second $^1\Pi_g$ state relative to the energy of the 2s$^2$2p$^2$($^3P$)+2s$^2$2p$^2$($^3P$) separated-atom limit.

Note that a global multi-state spectroscopic fit model for the $^1\Pi_g$ states is not currently possible, because the available experimental data~\cite{Dieke1930,Herzberg1940,Phillips1950,Messerle1967} only sample the energy region up to the lowest-energy curve-crossing between the d$_1$ and d$_2$ states. Even for the effective two-state model, some of the fit parameters, in particular the $h$ parameter for the d$_2$ diabat, are not well determined, due to lack of experimental observations of vibrational levels that belong to the second $^1\Pi_g$ electronic state. We find that use of a larger $h$ value for the shallow d$_2$ potential leads to a better fit to the observed $C$-state level energies. However, when this $h$ value is too large ($>$250), the inner well of the second $^1\Pi_g$ adiabatic potential from the fit model (orange curve in Fig.~\ref{fig:vh_two_state_fit}) bends down significantly relative to the DMRG result. In our fit model, the $h$ parameter for the d$_2$ diabat is fixed  at 225 to maintain a qualitative agreement between our fit model and the $ab$ $initio$ results at the inner well of the second adiabat. The $h$ parameter for the d$_1$ valence-hole diabat is fixed to the same value as derived from the global diabatic deperturbation of the $^1\Pi_g$ potentials from the DMRG calculation (Table~\ref{tab:valence_hole_1Pig}).

\begin{figure}
\center
\includegraphics[width=3 in]{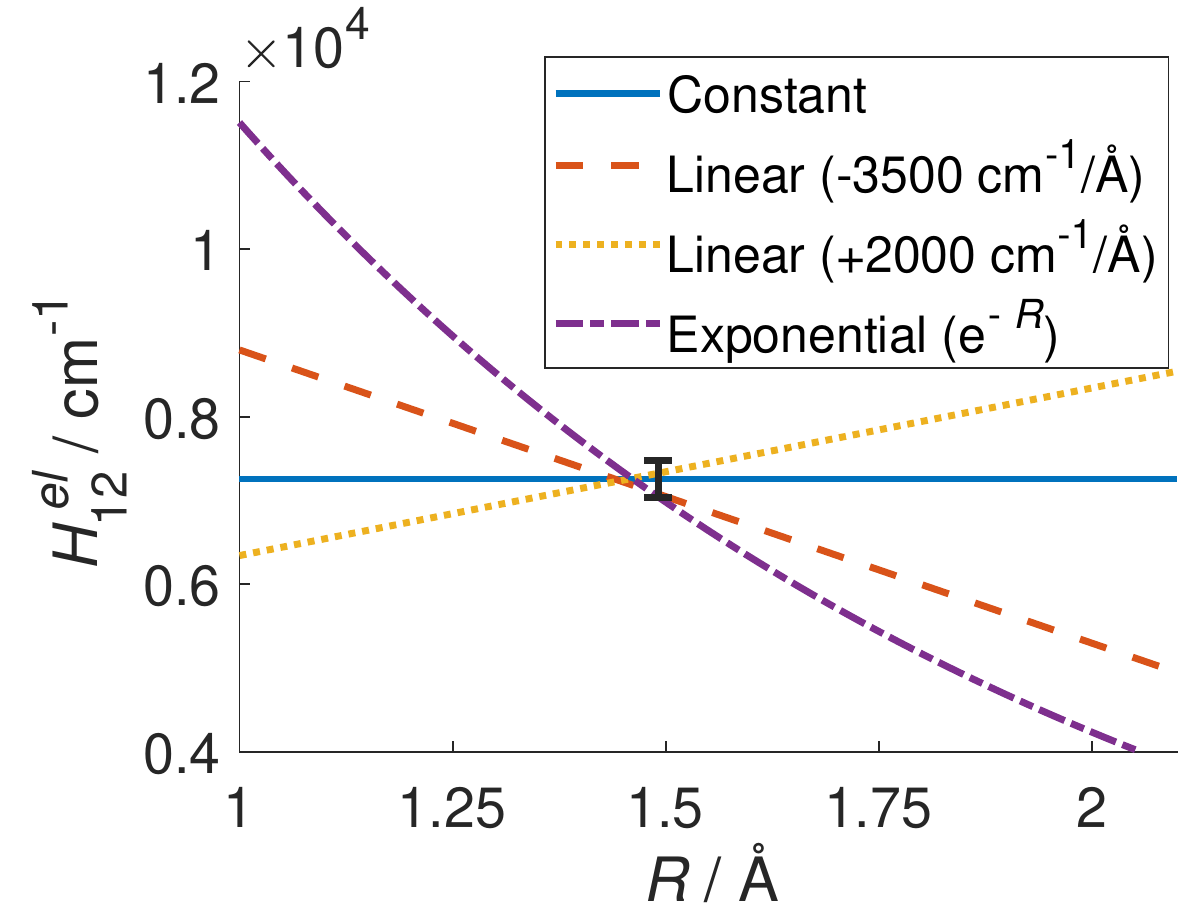}
\caption{The derived electronic matrix element, $H_{12}^{el}$, as a function of $R$, for the four tested functional forms of $H_{12}^{el}(R)$ used in the valence-hole two-state fit model. The error bar on the figure indicates the standard deviation of the $R$-independent electronic matrix element derived from a fit that assumes a constant $H_{12}^{el}$ value (Table~\ref{tab:valence_hole_two_state}).}
\label{fig:R_dependence}
\end{figure}

The effect of the choice of the functional form of the electronic matrix element, $H_{12}^{el}(R)$, on the results of the two-state fit model is investigated with four different types of $R$-dependent $H_{12}^{el}$ terms: constant, linearly-increasing, linearly-decreasing, and exponential-decay. Despite the qualitatively different functional forms of $H_{12}^{el}(R)$ that are tested with the valence-hole two-state model, only modest changes are observed in the numerical accuracy of the four resulting fits. Compared to the average fit residual (0.4 cm$^{-1}$) of the two-state model with a constant $H_{12}^{el}$ term (discussed earlier in this section), the average fit residuals are both about 0.1 cm$^{-1}$ smaller for the two models for which the $H_{12}^{el}$ term decreases as $R$ increases (i.e. linearly-decreasing and exponential-decay). The fit is slightly worse for the model that assumes a linearly-increasing $H_{12}^{el}(R)$ matrix element (0.5 cm$^{-1}$ average fit residual).

The derived electronic matrix elements as a function of $R$ for all four tested functional forms of $H_{12}^{el}$ are shown in Fig.~\ref{fig:R_dependence}. As is evident from Fig.~\ref{fig:R_dependence}, all four $H_{12}^{el}$ matrix elements converge to the same value at the curve-crossing region of the two diabats ($R\sim1.49\,\textup{\AA}$). This result, as well as the observed modest impact on the fit accuracy from the functional form of $H_{12}^{el}(R)$, both reflect the fact that the electrostatic interaction matrix elements, $H_{1,v_1,J;2,v_2,J}$ (Eq.~2), are sampled most strongly at the $R$-value of the intersection between the two diabats.

\section{CASSCF calculation for the $^{1,3}\Pi_g$ states}
\label{sec:fifthroot}

\begin{figure}[t]
\includegraphics[width=6.3 in]{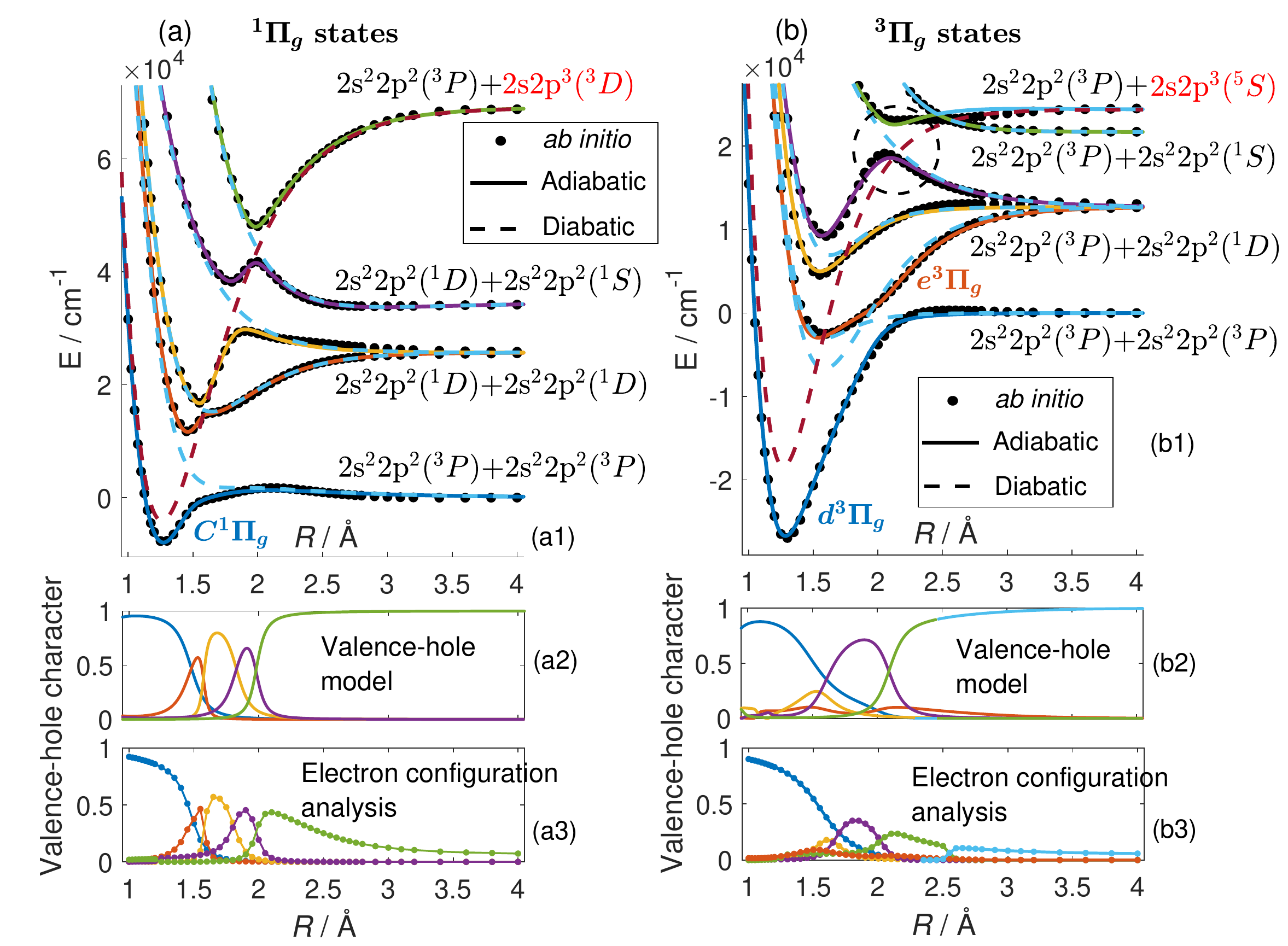}
\caption{The $ab$ $initio$ results, calculated with the CASSCF method, for the low-lying $^{1}\Pi_g$ and $^{3}\Pi_g$ states. Our global diabatization schemes, based on the valence-hole concept, for the electronic structure of the two symmetry species are shown in panels (a1) and (b1). The valence-hole character from the valence-hole model and the electron configuration analysis based on the CASSCF calculation are shown, respectively, in the lower two panels of each column, column (a) for the $^{1}\Pi_g$ states and column (b) for the $^{3}\Pi_g$ states.}
\label{fig:pi_valence_hole}
\end{figure}

The $ab$ $initio$ potentials for the $^{1,3}\Pi_g$ states, calculated by the CASSCF method (see Section IIIB of the main text for details), are shown in Fig.~\ref{fig:pi_valence_hole}. We have obtained more roots at each $R$-value using the CASSCF method than with the high-level DMRG (for the $^1\Pi_g$ states) and MRCI (for the $^3\Pi_g$ states) approach. The CASSCF calculation is used to confirm the validity of our global diabatic model of the $^{1,3}\Pi_g$ states, in particular, regarding the highest-energy $^{1,3}\Pi_g$ states which are not calculated by these high-level methods.


The results from the CASSCF calculation for the $^1\Pi_g$ states (top panel of Fig.~\ref{fig:pi_valence_hole}a) confirms the qualitatively correct shape for the fifth $^1\Pi_g$ adiabatic potential in our global diabatic interaction model of the $^1\Pi_g$ electronic structure (see Fig.\,4 of the main text). The pattern of transfer of valence-hole character across consecutive $^1\Pi_g$ states predicted by our valence-hole model (Fig.~\ref{fig:pi_valence_hole}a2) is consistent with the CASSCF electron configuration analysis in Fig.~\ref{fig:pi_valence_hole}a3, which shows the contribution from the valence-hole configuration ($2\sigma_g^22\sigma_u^11\pi_{u}^33\sigma_g^2$) as a function of $R$ in the lowest five $^1\Pi_g$ states.

The CASSCF calculation for the $^3\Pi_g$ states (top panel of Fig.~\ref{fig:pi_valence_hole}b) indicates that the diabatic state that converges to the 2s$^2$2p$^2$($^3P$)+2s$^2$2p$^2$($^1S$) limit is repulsive, and interacts weakly with the $^3\Pi_g$ valence-hole state (dashed brown). This interaction is neglected in our global diabatic model for the $^3\Pi_g$ states (see Fig.~5 of the main text). Note that the calculated energy of the 2s$^2$2p$^2$($^3P$)+2s2p$^3$($^5S$) fragment channel relative to the ground-state, 2s$^2$2p$^2$($^3P$)+2s$^2$2p$^2$($^3P$) fragment channel is lower by $\sim$1.25\,eV than the experimental value. The calculated energy of 2s$^2$2p$^2$($^3P$)+2s2p$^3$($^5S$) is only slightly higher than the energy of an avoided-crossing between the fourth and fifth $^3\Pi_g$ potentials (indicated by the dashed circle in Fig.~\ref{fig:pi_valence_hole}b1). As a result, the fifth $^3\Pi_g$ potential from the CASSCF calculation has a relatively flat trajectory as it follows this avoided-crossing.

\section{Numerical details of various global diabatic models}
\label{sec:numericals}

\subsection{Global deperturbation of the $^{1,3}\Pi_g$ states}
\label{sec:pig_states}

The molecular parameters for the diabatic states used in the global diabatic deperturbation of the low-lying $^1\Pi_g$ (Fig. 4 of the main text) and $^3\Pi_g$ (Fig. 5 of the main text) states are listed, respectively, in Tables~\ref{tab:valence_hole_1Pig} and \ref{tab:valence_hole_3Pig}. The diabatic state parameters for the $^1\Pi_g$ states are determined from a fit to the $ab$ $initio$ potentials calculated by the high-level DMRG method. The construction of the global fit model for the $^3\Pi_g$ states requires additional inputs from the MRCI calculation, as explained below.

Due to the close proximity in both the energies and $R$-values of the lowest three valence-hole-induced curve-crossings (indicated by the three arrows in Fig.~5a of the main text), the diabatic interactions among the low-lying $^3\Pi_g$ states are more complex in the energy region close to the 2s$^2$2p$^2$($^3P$)+2s$^2$2p$^2$($^3P$) dissociation threshold than those in the $^1\Pi_g$ manifold. The diabatic interactions in the $^3\Pi_g$ manifold are further complicated by an additional curve-crossing between the lowest two $normal$ valence states in the dashed boxed region of Fig.~5a. In our construction of the global diabatic interaction model for the $^3\Pi_g$ potentials from the MRCI calculation, we find it necessary to include the non-adiabatic interactions from the MRCI calculation as inputs to the fit model, in order to more accurately reproduce both the maximum values and the strong $R$-dependence of these non-adiabatic interactions. The non-adiabatic interactions between three pairs of neighboring $^3\Pi_g$ states from the MRCI calculation ($d$-$e$, $e$-3, and 3-4) are included in the fit.

In Tables~\ref{tab:valence_hole_1Pig} and \ref{tab:valence_hole_3Pig}, the energy of the 2s$^2$2p$^2$($^3P$)+2s$^2$2p$^2$($^3P$) limit is taken to be zero. The ``$T_e$\,(Diss)'' column gives the energies of the minima of various diabatic potential curves relative to the energy of that dissociation limit. In the last column of Tables~\ref{tab:valence_hole_1Pig} and \ref{tab:valence_hole_3Pig}, the 2s$^2$2p$^2$ label is omitted for all the atomic LS states with that electron configuration. 


\begin{table} 
\caption{Molecular parameters used in the global diabatization scheme for the $^1\Pi_g$ states calculated by the DMRG method. The exponential decay rates ($s_{ij}$) for all four pairs of electrostatic interactions ($\mathcal{H}_{ij}e^{-s_{ij} R}$) are assumed to be the same ($s_{ij}=s=1\,\text{\AA}^{-1}$). }
\begin{center}
\begin{tabular}{@{\hspace{8pt}} c @{\hspace{8pt}} | @{\hspace{6pt}} c  @{\hspace{6pt}} | @{\hspace{6pt}} c   @{\hspace{6pt}}| c  @{\hspace{6pt}} c   @{\hspace{6pt}} | @{\hspace{6pt}} c   @{\hspace{6pt}}| c   @{\hspace{6pt}} c @{\hspace{6pt}} @{\hspace{6pt}} c   @{\hspace{6pt}}c  | @{\hspace{6pt}} c @{\hspace{6pt}} | @{\hspace{6pt}} c @{\hspace{6pt}}}

\hline
diabats & $T_e$\,(Diss)\,/\,cm$^{-1}$ & $\beta$\,/\,$\text{\AA}^{-1}$ & & $R_e\,/\,\text{\AA}$& $h$ & Diss. Limit\\
\hline
d$_1$& -10871 & 2.2380 & &	1.2502 & 0.726 & ($^3P$)+2s2p$^3$($^3D$)\\
d$_2$& -3245 & 3.8895 & & 1.6139 & 40.5 & ($^3P$)+($^3P$) \\
d$_3$& 9245 & 2.3185 & & 1.6306 & 2.23 & ($^1D$)+($^1D$) \\
d$_5$& 29745 & 2.3221 & & 2.3146 & 0 (fixed) & ($^1D$)+($^1S$) \\
\hline
 & $\mathcal{A}_r$\,/\,cm$^{-1}$ & $k_r$\,/\,$\text{\AA}^{-1}$ & & &  & \\
\hline
d$_4$ & 3188300 & 3.7596 & &	&  & ($^1D$)+($^1D$)\\
\hline\hline
 interaction & $\mathcal{H}_{ij}$\,/\,cm$^{-1}$ & interaction && $\mathcal{H}_{ij}$\,/\,cm$^{-1}$ & &\\
\hline
d$_1$-d$_2$ & 29764 & d$_1$-d$_4$&& 30727 &&\\
d$_1$-d$_3$ & 7040 & d$_1$-d$_5$&& 27830 &&\\

\end{tabular}
\end{center}
\label{tab:valence_hole_1Pig}
\end{table}

\begin{table} 
\caption{Molecular parameters used in the global diabatization scheme for the $^3\Pi_g$ states calculated by the MRCI method. The exponential decay rates ($s_{ij}$) for all five pairs of electrostatic interactions ($\mathcal{H}_{ij}e^{-s_{ij} R}$) are assumed to be the same. These decay rates are determined to be $s_{ij}=s=0.587\,\text{\AA}^{-1}$ from the fit to the MRCI results.}
\begin{center}
\begin{tabular}{@{\hspace{8pt}} c @{\hspace{8pt}} | @{\hspace{6pt}} c  @{\hspace{6pt}} | @{\hspace{6pt}} c   @{\hspace{6pt}}| c  @{\hspace{6pt}} c   @{\hspace{6pt}} | @{\hspace{6pt}} c   @{\hspace{6pt}}| c   @{\hspace{6pt}} c @{\hspace{6pt}} @{\hspace{6pt}} c   @{\hspace{6pt}}c  | @{\hspace{6pt}} c @{\hspace{6pt}} | @{\hspace{6pt}} c @{\hspace{6pt}}}

\hline
diabats & $T_e$\,(Diss)\,/\,cm$^{-1}$ & $\beta$\,/\,$\text{\AA}^{-1}$ & & $R_e\,/\,\text{\AA}$& $h$ & Diss. Limit\\
\hline
d$_1$& -28619 & 2.6916 & &	1.2563 & 4.11 & ($^3P$)+2s2p$^3$($^5S$)\\
d$_2$& -10188 & 2.4015 & & 1.5710 & 4.14 & ($^3P$)+($^1D$) \\
d$_3$& -7747 & 3.2301 & & 1.5997 & 18.1 & ($^3P$)+($^3P$) \\
d$_4$& 1188 & 3.3329 & & 1.6039 & 32.2 & ($^3P$)+($^1D$) \\
\hline
 & $\mathcal{A}_r$\,/\,cm$^{-1}$ & $k_r$\,/\,$\text{\AA}^{-1}$ & & &  & \\
\hline
d$_5$ & 5576900 & 3.1907 & &	&  & ($^3P$)+($^1D$)\\
\hline\hline
 interaction & $\mathcal{H}_{ij}$\,/\,cm$^{-1}$ & interaction && $\mathcal{H}_{ij}$\,/\,cm$^{-1}$ & interaction & $\mathcal{H}_{ij}$\,/\,cm$^{-1}$\\
\hline
d$_1$-d$_2$ & 19694 & d$_1$-d$_4$&& 6924 &d$_2$-d$_3$&8006\\
d$_1$-d$_3$ & 2695 & d$_1$-d$_5$&& 14411 &&\\

\end{tabular}
\end{center}
\label{tab:valence_hole_3Pig}
\end{table}

\newpage
\subsection{Global deperturbation for the $^{1,3}\Sigma_u^+$ states}
\label{sec:sigmau_states}

The molecular parameters for the diabatic states used in the global diabatic deperturbation of the low-lying $^1\Sigma_u^+$ (Fig.~6a) and $^3\Sigma_u^+$ states (Fig.~6b) are listed, respectively, in Tables~\ref{tab:valence_hole_1sigmau} and \ref{tab:valence_hole_3sigmau}. These parameters are determined from a fit to the CASSCF $ab$ $initio$ potentials.

\begin{table}
\caption{Molecular parameters used in the global diabatization scheme for the $^1\Sigma_u^+$ states calculated by the CASSCF method. The energy of the 2s$^2$2p$^2$($^1D$)+2s$^2$2p$^2$($^1S$) limit is taken to be zero. The exponential decay rate ($s_{12}$) of the electrostatic interaction ($\mathcal{H}_{12}e^{-s_{12} R}$) is fixed to $s_{12}=1\,\text{\AA}^{-1}$. In the last column, the 2s$^2$2p$^2$ label is omitted for all the atomic LS states with that electron configuration.}
\begin{center}
\begin{tabular}{@{\hspace{8pt}} c @{\hspace{8pt}} | @{\hspace{6pt}} c  @{\hspace{6pt}} | @{\hspace{6pt}} c   @{\hspace{6pt}}| c  @{\hspace{6pt}} c   @{\hspace{6pt}} | @{\hspace{6pt}} c   @{\hspace{6pt}}| c   @{\hspace{6pt}} c @{\hspace{6pt}} @{\hspace{6pt}} c   @{\hspace{6pt}}c  | @{\hspace{6pt}} c @{\hspace{6pt}} | @{\hspace{6pt}} c @{\hspace{6pt}}}

\hline
diabats & $T_e$\,(Diss)\,/\,cm$^{-1}$ & $\beta$\,/\,$\text{\AA}^{-1}$ & & $R_e\,/\,\text{\AA}$& $h$ & Diss. Limit\\
\hline
d$_1$& -29360 & 2.0977 & &	1.2738 & 0.370 & ($^3P$)+2s2p$^3$($^3D$)\\
d$_2$& -3611 & 1.8814 & & 2.0823 & 1.62& ($^1D$)+($^1S$) \\
\hline\hline
 interaction & $\mathcal{H}_{12}$\,/\,cm$^{-1}$ & $s_{12}$\,/\,$\text{\AA}^{-1}$ && & &\\
\hline
d$_1$-d$_2$ & 10256 & 1*&&  &&\\

\end{tabular}
\end{center}
\label{tab:valence_hole_1sigmau}
\end{table}

\begin{table}
\caption{Molecular parameters used in the global diabatization scheme for the $^3\Sigma_u^+$ states calculated by the CASSCF method. The energy of the 2s$^2$2p$^2$($^3P$)+2s$^2$2p$^2$($^3P$) limit is taken to be zero. The exponential decay rates ($s_{ij}$) for all three pairs of electrostatic interactions ($\mathcal{H}_{ij}e^{-s_{ij} R}$) are assumed to be the same. These decay rates are determined to be $s_{ij}=s=0.882\,\text{\AA}^{-1}$ from the fit to the CASSCF potentials. In the last column, the 2s$^2$2p$^2$ label is omitted for all the atomic LS states with that electron configuration.}
\begin{center}
\begin{tabular}{@{\hspace{8pt}} c @{\hspace{8pt}} | @{\hspace{6pt}} c  @{\hspace{6pt}} | @{\hspace{6pt}} c   @{\hspace{6pt}}| c  @{\hspace{6pt}} c   @{\hspace{6pt}} | @{\hspace{6pt}} c   @{\hspace{6pt}}| c   @{\hspace{6pt}} c @{\hspace{6pt}} @{\hspace{6pt}} c   @{\hspace{6pt}}c  | @{\hspace{6pt}} c @{\hspace{6pt}} | @{\hspace{6pt}} c @{\hspace{6pt}}}

\hline
diabats & $T_e$\,(Diss)\,/\,cm$^{-1}$ & $\beta$\,/\,$\text{\AA}^{-1}$ & & $R_e\,/\,\text{\AA}$& $h$ & Diss. Limit\\
\hline
d$_1$& -33379 & 3.6848 & &	1.2400 & 37.5 & ($^3P$)+2s2p$^3$($^5S$)\\
d$_2$& -10188 & 4.6108 & & 1.4365 & 154 & ($^3P$)+($^3P$) \\
d$_3$& -2917 & 4.3827 & & 1.7297 & 295 & ($^3P$)+($^3P$) \\
\hline
 & $\mathcal{A}_r$\,/\,cm$^{-1}$ & $k_r$\,/\,$\text{\AA}^{-1}$ & & &  & \\
\hline
d$_4$ & 6363500 & 3.0599 & &	&  & ($^3P$)+($^1D$)\\
\hline\hline
 interaction & $\mathcal{H}_{ij}$\,/\,cm$^{-1}$ & interaction && $\mathcal{H}_{ij}$\,/\,cm$^{-1}$ & interaction & $\mathcal{H}_{ij}$\,/\,cm$^{-1}$\\
\hline
d$_1$-d$_2$ & 39979 & d$_1$-d$_3$&& 5736 &d$_1$-d$_4$&12873\\

\end{tabular}
\end{center}
\label{tab:valence_hole_3sigmau}
\end{table}

\bibliography{C2_JPCA_supplemental}